\documentclass[twocolumn]{autart}

\usepackage{graphicx}       
\usepackage{amsmath,amsfonts,amssymb}
\usepackage{xcolor}
\usepackage{url}
\usepackage{siunitx}

\usepackage{algpseudocode}  

\usepackage{fancyvrb}   
\usepackage{listings}

\newcommand{\zono}[1]{{\langle{#1}\rangle}}
\newcommand{\redop}[1]{{\downarrow_{#1}}}
\newcommand{\rand}[1]{\mathbf{#1}}
\newcommand{\func}[1]{\mathrm{#1}}

\newcommand{\inter}[0]{\square}   
\newcommand{\sign}[0]{|\!_-^+\!|}
\newcommand{\bool}[0]{|_0^1|}
\newcommand{\imset}[1]{{\langle{#1}\rangle}}
\newcommand{\identity}{\mathcal{I}}

\newcommand{\nnd}[0]{\barwedge}    
\newcommand{\eqa}[0]{\leftarrow}   

\newcommand{\bmid}[1]{\hat{#1}}
\newcommand{\brad}[1]{\mathring{#1}}
\newcommand{\binf}[1]{\underline{#1}}
\newcommand{\bsup}[1]{\overline{#1}}

\newcommand{\spoly}[1]{\mathbf{#1}}

\newcommand{\dom}[1]{{\mathcal{D}{#1}}}

\newcommand{\llbracket}[0]{[\![}
\newcommand{\rrbracket}[0]{]\!]}

\newcommand{\token}[1]{\mathtt{<}\mathtt{#1}\mathtt{>}}

\newcommand{\REVo}[1]{#1}

\begin{document}

\begin{frontmatter}

\title{Functional sets with typed symbols : Mixed zonotopes and Polynotopes for hybrid nonlinear reachability and filtering}

\thanks[footnoteinfo]{This paper was not presented at any IFAC meeting. Corresponding author: C.~Combastel. \copyright 2022. This manuscript version is made available under the CC-BY-NC-ND 4.0 license {https://creativecommons.org/licenses/by-nc-nd/4.0/}}

\author[UB]{C. Combastel}{\ead{christophe.combastel@u-bordeaux.fr}}     
\address[UB]{Univ. Bordeaux, CNRS, IMS, UMR 5218, 33405 Talence, France}


\begin{abstract}                          
Verification and synthesis of Cyber-Physical Systems (CPS) are challenging and still raise numerous issues so far. In this paper, based on a new concept of mixed sets defined as function images of symbol type domains, a compositional approach combining eager and lazy evaluations is proposed. Syntax and semantics are explicitly distinguished. Both continuous (interval) and discrete (signed, boolean) symbol types are used to model dependencies through linear and polynomial functions, so leading to mixed zonotopic and polynotopic sets. Polynotopes extend sparse polynomial zonotopes with typed symbols. Polynotopes can both propagate a mixed encoding of intervals and describe the behavior of logic gates. A functional completeness result is given, as well as an inclusion method for elementary nonlinear and switching functions. A Polynotopic Kalman Filter (PKF) is then proposed as a hybrid nonlinear extension of Zonotopic Kalman Filters (ZKF). Bridges with a stochastic uncertainty paradigm are briefly outlined. Finally, several discrete, continuous and hybrid numerical examples including comparisons illustrate the effectiveness of the theoretical results.
\end{abstract}

\end{frontmatter}

\section{Introduction} \label{sec:intro}

Uncertainty management undoubtedly remains a great challenge when designing, observing, controlling and verifying
systems with stringent safety, reliability and accuracy requirements. 
Common industrial practice still makes intensive use of Monte-Carlo simulations
and design of experiments
to check for robustness, perform sensitivity analysis and optimize tuning. Meanwhile, given some model of the available knowledge, which is by essence subject to uncertainties, formal methods
provide verification and synthesis tools likely to ensure a full coverage wrt to the range of specified behaviors, including off-nominal and worst cases. Dealing with complex dynamics such as nonlinear and hybrid ones remains challenging and the achieved trade-off between computation time and accuracy heavily depends on underlying set representations. 
\\   
However, a direct use of set-membership techniques (e.g. intervals, ellipsoids, etc) is often subject to the so-called dependency problem and/or the wrapping effect. The former comes from the loss of variable multi-occurrences when overloading basic operators. The latter results from the necessary approximate (usually outer) description of true solution sets.
This has motivated the use of affine arithmetic \cite{Figueiredo2004,Rump2015} and other set representations of intermediate complexity between intervals \cite{Jaulin2001,Raissi2018,Dinh2020} or ellipsoids \cite{Kurzhanskiy2007} and polytopes \cite{Gruenbaum2003} or level sets \cite{Mitchell2008}, like zonotopes \cite{Kuhn1998,Combastel2003,Goubault2015}, a class of convex and centrally symmetric polytopic sets defined as the affine image of a unit hypercube. 
Zonotopes have been used to address the reachability of linear \cite{Puig2003,Girard2005,Combastel2011}, nonlinear \cite{Althoff2013,Althoff2014,Goubault2020} and hybrid e.g. \cite{Maiga2016} systems, as well as state bounding observation \cite{Combastel2003,Alamo2005,Combastel2005,Le2013}, possibly with links to a stochastic paradigm \cite{Combastel2015,Combastel2016}, or in a distributed context \cite{Orihuela2018,Combastel2019}. 
With zonotopes, affine function transforms correspond to implicit set operations and the evaluation of bounds can be delayed (lazy evaluation \cite{Sinot2008}), to the benefit of a better management of 
dependencies. 
\\  
Moreover, an analogy can be noticed between such affine function transforms and the manipulation of symbolic expressions at a syntactic level (e.g. $x-x$ simplified as $0$ before substituting the unit interval $[-1,+1]$ for $x$ in $x-x$). In addition, the important distinction between syntax (e.g. formal transformation rules) and semantics (e.g. the interpretation/evaluation of some expression) is easily lost when directly operating sets. 
By extending ideas originating from affine arithmetic \cite{Figueiredo2004} and further developed, e.g., for the static analysis of programs by abstract interpretation \cite{Goubault2015}, symbolic zonotopes and USP (Unique Symbol Provider) in \cite{Combastel2019} showed the relevance of these concepts in a context of distributed state estimation.
To summarize, a clear distinction between syntax and semantics is a key point to struggle against the dependency problem in set-membership computations. 

When addressing core problems like reachability, state estimation, identification, invariant sets \cite{Wang2019} and fault diagnosis \cite{Xu2015,Scott2016,Pourasghar2019}, non-convex and possibly non-connected sets are often required when dealing with nonlinear and/or hybrid systems. Taylor models \cite{Makino2009} are an alternative to represent non-convex sets. 
Dealing with non convex and/or non-connected sets is also possible through pavings \cite{Jaulin2001} or level sets \cite{Mitchell2008} but costly since related algorithms respectively rely on bisections or grids, both yielding an exponential complexity.
Using properties like monotony/cooperativity \cite{Ramdani2009} often impose restrictions either on the class of dynamics that can readily be handled, or on accuracy due to the lack of richer internal set descriptions. In the hybrid case, the crossing of guards can generate many pieces of flows. Bissection/branching may be used, but the benefit of fast methods is then often lost due to the complexity induced by the propagation of a large number of (possibly smaller) instances which unduly become fully independent
after branching/bissection. \\ 
Thus, two complementary directions can be considered and \emph{combined}:
finding more versatile set representations possibly 
$i)$ \emph{non convex} while preserving scalability, 
and 
$ii)$ \emph{non connected} to 
propagate unions/bundles of a possibly large number of (implicit) sets characterizing distinct/discrete configurations/modes, without full bissections/branching, that is, while sharing and keeping trace of the common features between all these sets.
The combination of $i)$ and $ii)$ pleads in favor of searching for some kind of unified representation to encode and operate \emph{mixed sets} (i.e. hybrid sets).	

Though zonotopic sets catch some linear dependencies, 
their convex, connected, and centrally symmetric nature still impose restrictions to address the reachability of nonlinear and hybrid (i.e. mixed continuous/discrete) dynamic systems. To overcome these restrictions, Taylor models \cite{Makino2009}, polynomial zonotopes \cite{Althoff2013} and sparse polynomial zonotopes (spz) \cite{Kochdumper2019} rely on sets defined as polynomial images rather than affine/linear ones. spz can thus efficiently store and operate a large class of non convex continuous sets. However, spz do not natively handle the case of discrete or mixed sets, which motivates the distinct features introduced with the polynotopes proposed in this paper. Moreover, constraints can be also introduced in set representations as with constrained zonotopes \cite{Scott2016} and, recently, constrained polynomial zonotopes \cite{Kochdumper2020}. Note that the evaluation of bounds under constraints, even if delayed, may be costly and involve iterative algorithms (e.g. linear programming with constrained zonotopes). Polynotopes will thus introduce typed symbols whose management involves specific polynomial constraints that can be efficiently handled.  

\emph{Contributions}. This paper introduces a new concept of jointly mathematical and computational objects called polynotopes allowing to define and operate functional sets with typed symbols. By focusing on one continuous (interval) and two discrete (signed, boolean) symbol types, it is shown how the resulting non convex, non centrally symmetric and non connected mixed sets extending zonotopes (including polynomial ones) can be used to implement advanced hybrid nonlinear reachability and filtering algorithms without relying on costly bissections. 
Syntax and semantics are explicitly distinguished while combining both a strict/eager evaluation of polynotope objects (mainly addressing the dependency problem) and a lazy/delayed evaluation of bounding sets (mainly addressing the wrapping effect). 
Mixed encoding of intervals and dependency preserving inclusion methods for elementary nonlinear and switching functions are proposed.
Polynomial representation of logic functions defined on $\{-1,+1\}$ (signed logic) or $\{0,1\}$ (boolean logic) is analyzed and a functional completeness result is given for polynotopes.
Based on operator overloading, the generic implementation of an original Polynotopic Kalman Filter (PKF) extending Zonotopic Kalman Filters (ZKF) to hybrid nonlinear systems is obtained.
Several discrete, continuous and hybrid numerical examples illustrate the main theoretical results.%

\emph{Organization.} After extending the notion of inclusion function classically used in interval arithmetic to general sets in section~\ref{sec:prelim}, the motivation and the construction/composition of polynotope objects is treated in section~\ref{sec:polynobjwhyhow}. Uniquely identified typed symbols are also introduced in this section. A possible implementation of the polynotope objects is then gradually introduced in section~\ref{sec:polynotopes} by first starting from symbolic/mixed zonotopes and mixed encoding of intervals, and then extending sparse polynomial zonotopes with specific features making it possible to handle mixed sets in a unified way. 
In section~\ref{sec:hybridtools}, modeling tools for nonlinear hybrid systems are given with emphasis placed on a compositional approach relying on basic logic gates and basic nonlinear continuous and switching functions.
Inclusion methods are also given.
Then, a Polynotopic Kalman Fiter (PKF) extending ZKF to hybrid nonlinear systems is developed in section~\ref{sec:filtering}. Through basic operators/functions overloading, its implementation can benefit from the proposed dependency preserving compositional inclusion methods. 
The links between PKF, ZKF and the basic stochastic Kalman Filter KF \cite{Kalman1960} are made explicit. 
In section~\ref{sec:numex}, numerical examples including comparisons illustrate the effectiveness of the proposed scheme, before concluding remarks in section~\ref{sec:concl}.

\section{Inclusion function: beyond intervals} \label{sec:prelim}

To begin with, a definition of sets from functions (imset) and a definition of inclusion functions are given and discussed in a classical (non-symbolic) framework.
Given a variable $x$, let $\dom{x}$ denote a set (or domain) of possible values for $x$, that is: $x\in\dom{x}$. Also, let $\dom{X}={^\subset\dom{x}}$ denote a set of subsets of $\dom{x}$, including $\dom{x}$ itself, that is: $(\forall X\in\dom{X},\, X\subset\dom{x})\wedge(\dom{x}\in\dom{X})$. Since the value of any $X\in\dom{X}$ is a set, $X$ is said \emph{set-valued}. Notice that $x \in \dom{x}$ is not necessarily set-valued.
For instance, the Table~\ref{table:domainexamples} reports some continuous, discrete and mixed examples, where $\bar{\mathbb{R}}=\mathbb{R}\cup\{-\infty,+\infty\}$,  
and $\bar{\mathbb{I}}X$ (resp. $\mathbb{I}X$) refers to the set (resp. collection) of real intervals $[a,b]$ included in some set $X$.
Thus, $\bar{\mathbb{I}}X=\{\mathbb{I}X\}$.
The continuous example with $\dom{X}=\bar{\mathbb{I}}\bar{\mathbb{R}}^n$ is classically used to define inclusion functions in the particular case of interval arithmetic with possibly unbounded\footnote{Note here that $\mathbb{R}^n=\,]{-\infty},{+\infty}[^n\,\in\dom{X}$, $\dom{x}=\bar{\mathbb{R}}^n=[-\infty,+\infty]^n\in\dom{X}$, and $\dom{X}=\bar{\mathbb{I}}\bar{\mathbb{R}}^n$ is a strict subset of the power set $2^\dom{x}$ of $\dom{x}=\bar{\mathbb{R}}^n$ (e.g. a sphere is not an interval).} intervals. 
\begin{table} 
	\caption{Examples of possible domains $\dom{x}$ and $\dom{X}$ for a 
		variable $x$} \label{table:domainexamples} 
	\begin{tabular}{lll}
		Example & $\dom{x}$ & $\dom{X}$ \\
		\hline
		\rule{0mm}{10pt}Continuous & 
		$\bar{\mathbb{R}}^n$ & $\bar{\mathbb{I}}\bar{\mathbb{R}}^n$\\
		Discrete & $\{0,1\}$ & $\{\{0\}, \{1\}, \{0,1\}\}$\\
		Mixed &  $\{0,1\}\cup[2,4]$ & $\{\{1\}, \mathbb{I}[2,3], \{0\}\cup[3,4], \dom{x}\}$\\
	\end{tabular}
\end{table}
\begin{defn}[imset] \label{defn:imset}
	Given a function $f:\dom{x}\rightarrow \dom{y},$ \\
	$x\mapsto y=f(x)$, and a set $X\subset\dom{x}$, the imset of $X$ by $f$ is: \\
	\rule{1cm}{0mm} $f(X)=\{f(x) \, | \, x\in X\}$.
\end{defn}
\begin{defn}[Inclusion function] \label{defn:inclfun}
	The function $g : \dom{X} \rightarrow \dom{Y},$ $X\mapsto Y=g(X)$ is an inclusion function for 
	$f:\dom{x}\rightarrow \dom{y},$ $x\mapsto y=f(x)$, if $\dom{x} \in \dom{X}$ and: \\
	\rule{1cm}{0mm} $\forall X \in \dom{X}$, $f(X) \subset g(X)$, \\
	where $f(X)$ is the imset of $X\subset\dom{x}$ by $f$, and $g(X)$ is the image of $X\in\dom{X}$ by $g$. 
\end{defn}
\begin{cor}	\label{cor:inclfuncor2}
	$g$ must not be monotone wrt inclusion to be an inclusion function. Even without this requirement, it can be inferred that: \\
	\rule{10mm}{0mm} $\forall X\in\dom{X}, \, f(X)\subset g(\dom{x})$.\\
\end{cor}
	\textbf{Proof.} 
	Firstly, $imset_f : \dom{X} \rightarrow \dom{Y},$ $X\mapsto Y=f(X)$ is monotone wrt inclusion: $X_1 \subset X_2 \,\Rightarrow\, imset_f(X_1) \subset imset_f(X_2)$.
	Moreover, from the definition~\ref{defn:inclfun}, $\forall X \in \dom{X}, \, X\subset\dom{x}$, and the monotony of $imset_f$ wrt inclusion gives: $\forall X\in\dom{X}, \, f(X) \subset f(\dom{x})$. Also, since $\dom{x} \in \dom{X}$, $f(\dom{x}) \subset g(\dom{x})$. Thus, $\forall X\in\dom{X}, \, f(X) \subset g(\dom{x})$, without requiring the additional statement that $g$ must be monotone wrt inclusion to be an inclusion function. 
\begin{rem}
	In general, $2^{\dom{x}} \not\subset \dom{X}$. 
	This may lead to some conservatism when using an inclusion function defined over $\dom{X}$ to enclose the image of any arbitrary subset $X\in2^{\dom{x}}$ of $\dom{x}$. 
	This kind of wrapping effect is hardly avoidable in practice since, in most cases, not all subsets of $\dom{x}$ can be exactly represented in machine, especially when dealing with continuous and/or hybrid domains.
\end{rem}
\begin{rem} 
	Syntax and semantics are not distinguished in this section~\ref{sec:prelim} where the same notation $x$ may refer to the (name of a) variable  or a taken value.
\end{rem}
\REVo{ 
A basic illustration of interval arithmetic's dependency problem is first given and analyzed. Let $x$ denote any real such that $x\in[-1,+1]\subset\mathbb{R}$. Let $f:\bar{\mathbb{R}}\rightarrow\bar{\mathbb{R}},\, x\mapsto x-x$.
$i)$ Since $\forall x,\, x-x=0$, $f$ is the null function and the imset $f([-1,+1])$ of the interval $[-1,+1]$ by $f$ is obviously the singleton $\{{0}\}$. 
$ii)$ Let $g$ be the natural interval extension of $f$ i.e. $g:\mathbb{I}\bar{\mathbb{R}}\rightarrow\mathbb{I}\bar{\mathbb{R}}, x\mapsto x-x$, where the minus operator is ``overloaded'' to deal with interval operands. $g$ is an inclusion function for $f$. Thus: $f([-1,+1])\subset g([-1,+1]) = [-1,+1]-[-1,+1]=[-2,+2]$.
Conclusion: Though inclusion is preserved, $[-2,+2]$ is a poor outer approximation of $f([-1,+1])=\{0\}$. 
} 
Whereas simply overloading basic \REVo{interval } operators would be the dream of a programmer wishing to implement uncertainty propagation computations, the natural interval extension is often subject to a significant conservatism. This is mainly due to the loss of dependencies between multiple occurrences of variables in the expressions to be evaluated (e.g. notice that $x$ occurs twice in $f$ and $g$). Indeed, important symbolic links are lost when substituting some interval value (semantics) for a given variable name/symbol (syntax) occurring in expressions/formula used to define mathematical functions and/or imsets as in definition~\ref{defn:imset}.
\REVo{ 
Moreover, iterative evaluations, e.g. to compute reachable sets of dynamical systems, often emphasize the drawbacks of a natural interval extension.
}
Then, should we definitely abandon the dream of encoding guaranteed (inclusion preserving), accurate and fast uncertainty propagation algorithms simply by overloading the operators used to define and program explicit mathematical functions? Beyond the natural interval arithmetic extension, how to tackle the dependency problem while dealing with possibly mixed (hybrid: continuous and discrete), non-convex and non-connected sets in a unified way?

\section{Polynotope objects: Why and how?} \label{sec:polynobjwhyhow} \label{sec:framework} 

	\subsection{Functional sets}   \label{subsec:FunSets}
\REVo{ 
Starting from intervals to represent bounding sets over continuous domains, explicit descriptions of more general sets is a possible direction to struggle against the dependency problem. Though general polytopes may look attractive to represent/approximate a large class of convex sets, the enumeration of their vertices and/or faces often remains poorly scalable. Paving and extensive use of contractions and bissections (branching) is another possible direction. However, bissections often restrict the scalability of accurate set approximations as the space dimension increases. Implicit rather than explicit set descriptions have also received a significant attention along the years e.g. ellipsoids described through a symmetric and positive definite (spd) matrix or as the imset of a unit hypersphere by a linear function, zonotopes (resp. polynomial zonotopes) most often viewed/defined as the } imset of a unit hypercube  $[-1,+1]^p\subset\mathbb{R}^p$ by an affine (resp. polynomial) function $f:\mathbb{R}^p\rightarrow\mathbb{R}^n$. For instance, the zonotope $\zono{c,R}=\{c+Rs, \, s\in[-1,+1]^p\}\subset\mathbb{R}^n$ is nothing else but the imset of $[-1,+1]^p$ by the function $f_{c,R}:s\mapsto c+Rs$. Though the possible shapes of zonotopes in $\mathbb{R}^n$ with $n>2$ are significantly more versatile than that of basic $n$D intervals, so usually providing very useful degrees of freedom to struggle against the wrapping effect, it remains worth noticing the consequences of the intrinsic symbolic (syntactic) \emph{difference} between the two expressions $f_{c,R}(s_a)$ and $f_{c,R}(s_b)$ where, e.g., $s_a\in[-1,+1]^p$ and $s_b\in[-1,+1]^p$. First, the possible values (semantics) taken by both expressions belong to the \emph{same} zonotope $\zono{c,R}$.
\REVo{ 
Indeed, the set-based wrap/enclosure $[f_{c,R}(s_a)]$ of $f_{c,R}(s_a)$ is $[f_{c,R}(s_a)]=f_{c,R}([s_a])=f_{c,R}([-1,+1]^p)=\zono{c,R}$ and idem with $s_b$ since $[s_b]=[s_a]=[-1,+1]^p$ (and so, even if $s_a \neq s_b$). In this context, what about the composition of set-based enclosures? \\
\rule{0pt}{12pt}$[f_{c,R}(s_a)+f_{0,-R}(s_a)]$ \ldots \\ 
$a_1) \quad \ldots\subset [c+Rs_a-Rs_a] \subset \{c\} = \zono{c,0}$ \\
$a_2) \quad \ldots\subset [f_{c,R}(s_a)]+[f_{0,-R}(s_a)] \subset \zono{c,2R}$ \\
\rule{0pt}{12pt}$[f_{c,R}(s_b)+f_{0,-R}(s_a)]$ \ldots \\
$b_1) \quad\ldots\subset [c+Rs_b-Rs_a] \subset \{c\}+R[s_b]-R[s_a] \subset \zono{c,2R}$ \\
$b_2) \quad \ldots\subset [f_{c,R}(s_b)]+[f_{0,-R}(s_a)] \subset \zono{c,2R}$ \\
\rule{0pt}{12pt}In the cases $a_1$ and $b_1$, a transformation/simplification of the symbolic expressions is first conducted (possibly based on an eager evaluation of terms interpreted as symbols rather than values) and the evaluation of set-based enclosures is delayed as much as possible (lazy evaluation).
In the cases $a_2$ and $b_2$, an eager evaluation using set-based enclosures as possible values is performed without any prior symbolic transformation. 
} 
In the case of $a_2$ and compared to $a_1$, this second approach yields a significant conservatism originating from the loss of dependencies between the two (thus, multiple) occurrences of $s_a$ which distinguish $a_1,a_2$ from $b_1,b_2$.

At this step, several directions orienting the following of this work can be drawn:\\
$i)$ Functions can be used to implicitly define sets as the image of their definition domain (e.g. unit hypercubes with affine functions $f_{c,R}$ for classical zonotopes).\\
$ii)$ Working out a traceability of variable multi-occurrences within the (possibly incrementally built) symbolic expressions used to define such functions/sets can help to struggle against the dependency problem.\\
$iii)$ While elementary operators overloading is preferred to encapsulate the code required to easily compose complex functions/sets from simple ones, a global\footnote{not only at a host level but also at a wider scale to also cover the network nodes of distributed systems like CPS.} scope should be maintained for symbolic variables to prevent from losing global dependencies.\\
$iv)$ A trade-off between the efficiency of symbolic operations, their memory footprint, and the ability to quickly compute guaranteed and accurate set-based enclosures has to be found, possibly involving reduction schemes.\\
$v)$ By default, binding/linking/sharing is preferred to systematic bissection/branching in this work. \\
$vi)$ Set-based enclosures should be computed only when needed (call by need), so leading to a lazy/delayed evaluation of wrapping sets and/or bounds. \\
$vii)$ The tight interactions between symbolic expressions and mathematical definitions as well as numerical computations reveals the need for distinguishing between syntax (symbolic level) and semantics (mathematical interpretations of symbolic expressions possibly resulting from numerical computations).

	\subsection{Syntax and semantics}   \label{subsec:SyntaxSemantics}

With the ultimate goal of better managing dependencies that constitute a key for an accurate propagation of uncertainties, an explicit distinction between syntax and semantics is considered.
Whereas syntax refers to rules defining symbol combinations that are correct in some language, semantics refers to the interpretation or meaning of related sentences. In other words, syntax refers to how writing correct statements, semantics indicates what they mean.

Let $z$ denote a (symbolic) name that should be understood/interpreted as an object $\iota z$. The interpretation operator $\iota$ assigns a meaning (e.g. mathematical or computational object, set, numerical value, etc) to the symbolic name $z$ (and possibly to other symbols/names). 
In this paper, $\iota$ is a generic notation used to disambiguate symbolic names (syntax) from their meaning and/or taken values (semantics), whenever necessary. 
$\iota z$ can be viewed as a shortcut for $\iota(z)$ or $\llbracket z\rrbracket_\iota$ which is a usual notation for valuation (semantics) in logic and computer science. However, systematic formal definitions of interpretation/valuation operators are out of scope in this paper: $\iota$ simply reads as ``interpretation of'' for the sole purpose of drawing an explicit separation between operations at syntactic and semantic levels.

For example, $\mathcal{D}x$ denoting a set containing the possible values (definition domain) of a variable named/symbolized by $x$. Then, $\iota x\in\mathcal{D}x$ reads as ``the interpretation/value assigned to the variable named $x$ belongs to the definition domain of $x$''. Notice that usual mathematical notations do not distinguish between variable names and values (e.g. when writing $x\in\mathcal{D}x$ for the variable $x$, as in section~\ref{sec:prelim}). 
Similarly, $f$ may denote the name/symbol referring to a function explicitly denoted as $\iota f$. Then, $\iota f$ stands for an interpretation of $f$ (e.g. as a mathematical function, as a subroutine/algorithm, etc). Moreover, an evaluation $\iota f(\iota x)$ of the function named $f$ stands for the image/result obtained by applying $\iota f$ to the value $\iota x$ assigned to the input variable named $x$.

	\subsection{Typed symbols and unique identifiers}   \label{subsec:TypedSymbols}

The so-called polynotope objects are interpreted (semantics) as image sets of vector polynomial functions defined from symbolic expressions (syntax) on domains related to different types of symbolic variables.
For instance, continuous (unit interval) and/or discrete (signed, boolean) symbols can be combined to define and operate possibly mixed/hybrid sets.
Following \cite{Combastel2019}, each of these symbols are uniquely identified in order to preserve dependencies while simplifying otherwise possible name binding issues.   
Considering other basic symbol types than interval ($\mathtt{i}$), signed ($\mathtt{s}$) and boolean ($\mathtt{b}$) is among possible extensions that are left to future works:
\begin{assum}[Basic types] \label{assum:types}
	Let $\mathbb{S}$ be a finite set of typed symbols. Each typed symbol $s_i\in\mathbb{S}$ is uniquely identified by an integer $i\in\mathbb{N}$.
	In this paper, the type $\tau s_i$ of any $s_i$ belongs to the set $\mathbb{T}=\{\mathtt{i}, \mathtt{s}, \mathtt{b}\}$ of three basic symbol types respectively referring to interval ($\mathtt{i}$), signed ($\mathtt{s}$) and boolean ($\mathtt{b}$).
	As shown in table~\ref{table:typedsymbols}, a definition domain is related to each of these basic types as $\inter=[-1,+1]$, $\sign=\{-1,+1\}$, $\bool=\{0,1\}$, respectively.
	By default, basic scalar values are assumed to be interpreted in the real field $\mathbb{R}$ equipped with the usual sum and product operators.
	A partition of $\mathbb{T}$ into continuous and discrete types is given by $\mathbb{T}=\mathbb{T}_c\cup\mathbb{T}_d$ with $\mathbb{T}_c=\{\mathtt{i}\}$ and $\mathbb{T}_d=\{\mathtt{s}, \mathtt{b}\}$.
\end{assum} 
\begin{defn}[Mixed, continuous, discrete] \label{defn:mcd}
	Let $I\subset \mathbb{N}$ and $T_I=\cup_{i\in I} \{\tau s_i\}$. The symbolic vector $s_I$ is mixed (resp. continuous, discrete) if $(T_I \cap \mathbb{T}_c \neq \emptyset) \wedge (T_I \cap \mathbb{T}_d \neq \emptyset)$ (resp. $T_I \subset \mathbb{T}_c$, $T_I \subset \mathbb{T}_d$). By extension, any formula $F(s_I)$ and/or related interpretation as (s-)function, (s-)zonotope, (s-)polynotope, etc can be qualified as mixed, continuous$\footnote{Regarding functions, continuous refers here to a property of the input domain and not to continuity as in analysis.}$ or discrete accordingly.
\end{defn}
\begin{cor} \label{cor:continuous}
	In an entirely continuous case, following the assumption~\ref{assum:types}, all the symbols in $s_I$ are of type (unit) interval: $\forall i\in I$, $\tau s_i=\mathtt{i}$, that is, $\forall i\in I$, $\iota \tau s_i=\inter=[-1;+1]$.
	As a result, for any vector $I$ of $n$ unique identifiers only referring to continuous symbols, any single-valued interpretation\footnote{as long as it is consistent with the type (unit) interval.} $\iota s_I$ of the symbolic vector $s_I$ belongs to the unit interval $\inter^n$, that is, $\iota s_I \in [-1;+1]^n$.
\end{cor}	
\begin{assum}[USP] \label{assum:USP}
	A Unique Symbol Provider (USP) is assumed to be implemented as a global function (or service in a distributed context) called as $!(n,t)$ with $(n,t)\in\mathbb{N}\times\mathbb{T}$ and then returning an $n$-dimensional vector $I\in\mathbb{N}^n$ of $n$ new unique identifiers referring to $n$ new typed symbols of type $t$ i.e. $\forall i\in I, \, \tau s_i=t$.
\end{assum}
\begin{rem}
	A simple implementation of $!(n,t)$ is ``$l=l+n$, $\func{return}(h(t)\mathbf{1}_n+2^{n_h}[l-n+1, ..., l])$'' where $\mathbf{1}_n$ denotes a vector of $n$ ones, $l$ is a persistent counter initialized to $0$ at startup, and $h:\mathbb{T} \rightarrow \mathbb{N}$ assigns to each type $t\in\mathbb{T}$ a unique integer tag encoded with at most $n_h$ bits. Under the assumption~\ref{assum:types}, reserving $n_h=2$ bits and taking $h(\{\mathtt{i}, \mathtt{s}, \mathtt{b}\})=\{1, 2, 3\}$ is a possible choice to efficiently manage type tags within the unique integer identifiers of typed symbols. Extensions include an overflow checking or an implementation (possibly distributed) as a service in a CPS (Cyber-Physical System): see \cite{Combastel2019} for details.	
\end{rem}

\begin{table}
	\caption{Notations for three basic symbol types in $\mathbb{T}=\{\mathtt{i}, \mathtt{s}, \mathtt{b}\}$ respectively referring to ``interval'', ``signed'', ``boolean''.}   \label{table:typedsymbols} 
	\begin{tabular}{lll}
		& syntax & semantics \\
		\hline
		typed symbol & $s_i=\mathtt{x}\mathtt{:}\mathtt{i}$ & $\iota s_i\in\mathcal{D}s_i$ \rule{12mm}{0mm}\emph{interval:} \\
		symbol type & $\tau s_i=\mathtt{i}$ & $\iota\tau s_i=\mathcal{D}s_i = \inter=[-1,+1]$ \\ 
		\hline
		typed symbol & $s_j=\mathtt{z}\mathtt{:}\mathtt{s}$ & $\iota s_j\in\mathcal{D}s_j$ \rule{12mm}{0mm}\emph{signed:} \\
		symbol type & $\tau s_j=\mathtt{s}$ & $\iota\tau s_j=\mathcal{D}s_j = \sign=\{-1,+1\}$ \\
		\hline
		typed symbol & $s_k=\mathtt{y}\mathtt{:}\mathtt{b}$ & $\iota s_k\in\mathcal{D}s_k$ \rule{12mm}{0mm}\emph{boolean:} \\
		symbol type & $\tau s_k=\mathtt{b}$ & $\iota\tau s_k=\mathcal{D}s_k = \bool=\{0,1\}$ \\ 
		\hline	
	\end{tabular}\\
	$\mathtt{x}$, $\mathtt{y}$, $\mathtt{z}$ : particular examples of (untyped) symbol names.\\ 
    The unique identifier of any typed symbol $s_i\in\mathbb{S}$ is $i\in\mathbb{N}$. 
\end{table}

	\subsection{Construction/composition of polynotope objects}   \label{subsec:BNF}

\begin{table}
\REVo{ 
	\caption{Context-free grammar in Backus-Naur Form (BNF) describing the syntax of a prototype language supporting the basic construction (through typed symbols) and composition of Polynotope objects.}   \label{table:BNF} 
	\includegraphics[width=84mm]{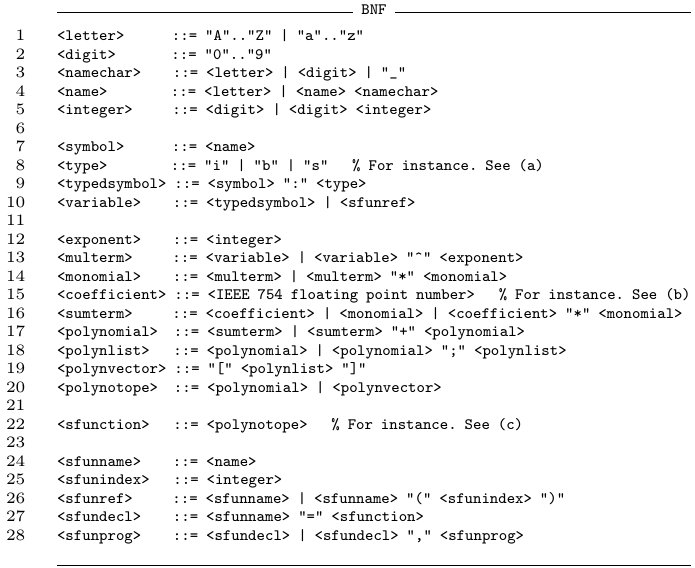}
} 
\tiny{%
\REVo{%
(a) Consistently with assumption~\ref{assum:types}. Polynotopes may support other types.\\
(b) Efficient numeric computations of constant/center vectors and coefficient/generator matrices of polynotope objects motivate this choice. \\
(c) At a syntactic level, canonical polynotopes (obtained after resolving \texttt{<sfunref>} references allowing to compose polynotopes) are polynomial vectors with only typed symbols as variables. Notice that vector linear functions (obtained from \texttt{<monomial> ::= <variable>} while removing \texttt{<exponent>}, \texttt{<multerm>}) give symbolic zonotopes and that nothing a priori hinders the definition of other \texttt{<sfunction>} as functions of typed symbols. } 
} 
\end{table}

\begin{table}
	\caption{Left: Sample code accepted by the grammar in Table~\ref{table:BNF}. Right: Evaluation of bounds resulting from a natural interval extension (int.) or some$^\dag$ polynotope computations (pol.).\label{table:samplecode}}
	\tiny$^\dag$~here, e.g., interval hull of an outer zonotope.\normalsize\\
\rule{0pt}{8pt}
\begin{center}
\rule{1mm}{0mm}
\begin{minipage}[c]{3.8cm}
\begin{Verbatim}[gobble=0,numbers=left,frame=lines,label={sample code},commandchars=\\\{\},commentchar=!,fontsize=\scriptsize]
u=symb:i,
x=0.5+0.5*u,
f1=[x*x; x],
f2=f1(2)-f1(1),
f3=[x^2; x],
f4=f1-f3,
r=remainder:i,
f5=[x+-0.125+0.125*r; x],
f6=f5(2)-f5(1)
\end{Verbatim}
\end{minipage}
\rule{1mm}{0mm}
\begin{minipage}[c]{1.6cm}
\begin{Verbatim}[gobble=0,frame=lines,label={int.},commandchars=\\\{\},commentchar=!,fontsize=\scriptsize]
[-1,+1]
[0,1] 
[0,1]^2
[-1,1]
[0,1]^2
[-1,+1]^2  
[-1,+1]
  (*)
[-1,5/4]
\end{Verbatim}
\end{minipage}
\rule{1mm}{0mm}
\begin{minipage}[c]{1.6cm}
\begin{Verbatim}[gobble=0,frame=lines,label={pol.},commandchars=\\\{\},commentchar=!,fontsize=\scriptsize]
[-1,+1]
[0,1]  
  (*)
[0,1/4]
  (*)
\{0\}^2
[-1,+1]
  (*)
[0,1/4]
\end{Verbatim}
\end{minipage}
\end{center}
\rule{0pt}{8pt}\rule{4.6cm}{0cm}\scriptsize{\texttt{(*) = [[-1/4,1];[0,1]]}}
\end{table}

Based on uniquely identified typed symbols as in \ref{subsec:TypedSymbols}, the remainder of this section~\ref{sec:polynobjwhyhow} aims at describing how the directions listed at the end of \ref{subsec:FunSets} can be achieved through polynotope objects, while introducing useful definitions to build well-grounded (e.g. inclusion preserving) mathematical interpretations. Consistently with the encapsulation principle \cite{Pierce2002}, internal data structures are left open in this section whereas some choice for them will be made explicit in section~\ref{sec:polynotopes}. \\
Polynotopes interprets as functional sets based on vector polynomial functions with typed symbols i.e. any polynotopic set can be viewed as the imset of the definition domain of a vector polynomial function depending on variables/symbols of different types such as, e.g., the basic types in assumption~\ref{assum:types}. Meanwhile, a mechanism compatible with operator overloading is needed to achieve the direction $iii)$ in \ref{subsec:FunSets}, that is, providing a simple interface for the user to naturally encode non trivial compositions of elementary polynotopes, while maintaining a) a global scope for typed symbols to struggle against the so-called dependency problem and b) the rigor of inclusion preserving set-based interpretations.  
\REVo{ 
For this purpose, a formal grammar describing the syntax of a prototype language is given in Table~\ref{table:BNF}. 
It supports a basic polynotope composition scheme (mainly based on the \texttt{<sfunref>} lexical tokens\footnote{A (lexical) token is a string (i.e. a sequence of characters) with an assigned and thus identified meaning.}), as exemplified with the sample code in Table~\ref{table:samplecode}.
Also, polynomials being closed under a finite\footnote{In this paper, a sufficiently large finite context is assumed. Concretely, this can be achieved through inclusion preserving reduction schemes, as in definition~\ref{defn:red} for instance.} number of compositions, it describes/accepts a set of well-formed formulas (wff) that can be operated at a symbolic level\footnote{e.g. by using efficient data structures to encode and manipulate the abstract syntax trees of vector polynomials with typed symbols: see (\ref{eq:spoly}) and symbolic polynotopes in \ref{subsec:polsfun}.}.
Such operations preserve the ability to delay as much as needed set-based evaluations of typed symbols (lazy evaluation), while providing the formal ground for inclusion preserving mathematical interpretations\footnote{e.g. see (\ref{eq:epoly}), e-polynotopes and definition~\ref{defn:rrincl} in \ref{subsec:polsfun}.}. 
Moreover, the eager evaluation of an \texttt{<sfunprog>} source code (like the sample in table~\ref{table:samplecode}) using polynotope objects allows to transform polynotopic symbolic function (s-function) declarations \texttt{<sfundecl>} (see also note (c) in Table~\ref{table:BNF}) into s-function definitions \texttt{<sfundef>}, that is, a canonical kind of \texttt{<sfunction>} such that all the references \texttt{<sfunref>} to other s-functions have been resolved\footnote{\texttt{<sfundef>} thus stands for an \texttt{<sfunction>} as in Table~\ref{table:BNF} except that \texttt{<sfunref>} is removed from line/rule~10 which then becomes \texttt{<variable>::=<typedsymbol>}.\label{ftn:sfundef}}. 
Concretely, polynotope objects store (the abstract syntax tree of) vector polynomial functions of typed symbols as sole variables, without any reference \texttt{<sfunref>} to other polynotope objects. \texttt{<sfunref>} references however remain a key enabler to compose non trivial polynotopes (by using intuitive operator overloading) from basic ones, which are typically constructed from \texttt{<typedsymbol>}s as shown at lines 1 and 7 in Table~\ref{table:samplecode}. 
Though an eager evaluation is firstly used to solve \texttt{<sfunref>} tokens \emph{only}, \texttt{<typedsymbol>} tokens are \emph{not} immediately evaluated. 
} 
This makes it possible to delay their evaluation while keeping trace of \emph{global} dependencies between polynotope objects since each typed symbol is encoded through a unique identifier. By tagging the symbol identifiers with their type, not only the symbols evaluation can be delayed (lazy evaluation) but also adapted to their type (polymorphism). As a result, polynotopes permit some kind of polymorphic delayed evaluation of uncertain symbols/variables. Moreover, a global scope is preserved for the latter, which is the basic key to address the dependency problem%
\REVo{ arising from a direct use of interval arithmetic. }%
The interplay between mathematical definitions and the semantics attached to the intermediate (polynotope objects) and finally computed values (e.g. zonotopes, intervals) also requires a special attention to ensure inclusion is preserved, not only on continuous domains, but also on discrete and mixed ones, as is made possible with polynotopes.

	\subsection{s-functions} \label{subsec:sfun}

In order to transform and evaluate expressions based on typed symbols, the notion of (polynotopic) symbolic function or, shortly, s-function, is precised. s-function definitions (sfd), some of their interpretations (sffi) and related evaluations (sffe) are then considered. Whereas sfd refers to syntax, sffi and sffe refer to semantics.

Let $\mathbb{F}$ denote the set of finite length well-formed formulas (wff) corresponding to canonical polynotopic symbolic functions as described by \texttt{<sfundef>} in paragraph~\ref{subsec:BNF} and, more specifically, in footnote~\ref{ftn:sfundef}. Let $f\in\mathbb{F}$ be such a wff. Firstly, $f$ describes a possibly scalar polynomial vector (see lines 20 and 22 in table~\ref{table:BNF}) of finite dimension $n_f\in\mathbb{N}$. Also, $f$ contains a finite number $p_f\in\mathbb{N}$ of typed symbols (\texttt{<typedsymbol>} tokens). Let $I=I_f\subset\mathbb{N}$ be the set\footnote{or a vector. Here, the ordering of scalar elements is free.} of their unique identifiers. Then, to emphasize its dependence on the typed symbols in $s_I$, the wff $f$ can be equivalently\footnote{Note that, alternately, $F(s_I)$ (resp. $F(.)$) may refer to the abstract syntax tree resulting from parsing the wff $f$ up to typed symbol leafs included (resp. not included).} denoted as $F(s_I)$ with $f=F(s_I)$.

\begin{defn}[s-function: sfd/sffi/sffe]:\\   \label{defn:sfun}
$\bullet$ s-function definition (sfd) : $f=F(s_I) \in \mathbb{F}$. $F(s_I)$ is a wff 
involving the (typed) symbolic variables in $s_I$ which become bound in the formula: indeed, $F(s_I)$ \emph{depends} on $s_I$, at least from a syntactical viewpoint. \\
$\bullet$ s-function functional interpretation (sffi) : denoted as $\iota f(.)$ to emphasize its \emph{functional nature}, an sffi of $f$ is an interpretation $\iota f$ of $f$ that has the \emph{ability} to define and/or return output values from input values corresponding to an interpretation/valuation of the (typed) symbols in $s_I$. \\ 
$\bullet$ s-function functional evaluation (sffe) : Let $\iota s_I$ be an interpretation/valuation of the symbolic vector $s_I$ and let $\iota f(.)$ be an sffi of $f=F(s_I)$. Then, the sffe $\iota f(\iota s_I)$ is the \emph{result} obtained by applying $\iota s_I$ to the sffi $\iota f(.)$.
\end{defn}
\REVo{ 
\begin{rem}[sffi vs sffe]
An s-function interpretation $\iota f$, as long as it is also an sffi $\iota f(.)$, should be understood as a mathematical function or subroutine or algorithm or any transformation process related to $f$ but \emph{not} as some returned output value obtained from such a function or process. 
By contrast, an sffe $\iota f(\iota s_I)$ should be understood as some \emph{returned value} obtained from an sffi $\iota f(.)$ fed by some interpretation/valuation $\iota s_I$ of the typed symbols $s_I$ referring to its (possibly uncertain) inputs. Though out of scope in this work, note that there is no incompatibility with taking functions as possible values, like in functional paradigms. Moreover, random variables being nothing else but functions from a set of outcomes to a set of possible values, the proposed approach is open to extensions involving stochastic descriptions.  
\end{rem}
} 

\REVo{ 
An s-function definition (sfd) usually results from the eager evaluation of some program containing s-functions declarations like, e.g., $\token{sfunprog}$ in Table~\ref{table:BNF}. For polynotopes, this evaluation is based on a systematic expansion leading to a polynomial vector with variables corresponding to typed symbols only, and no more reference to any other s-functions which are all resolved. In other words, the systematic polynomial expansion gives a fully unfolded abstract syntax tree, up to typed symbols i.e. s-function uncertain inputs.
For example, based on the sample code in lines 1-4 of Table~\ref{table:samplecode}, the following s-function definitions (sfd) are obtained: \\
\rule{4mm}{0mm}$f_1=$ \scriptsize\texttt{[0.25+0.5*symb:i+0.25*symb:i 2; 0.5+0.5*symb:i]}\normalsize, \\ 
\rule{4mm}{0mm}$f_2=$ \scriptsize\texttt{0.25-0.25*symb:i 2}\normalsize. \\ 
} 
\REVo{ 
Note that the elimination of \texttt{0.5*symb:i} in the sfd of $f_2$ is made possible at the symbolic level. This greatly helps to improve the evaluation of some bounds in Table~\ref{table:samplecode}. 
}

Several interpretations of an s-function $f=F(s_I)$ may coexist but a basic one is as a mathematical function like
\begin{align}   
	\iota f : \,\iota \tau s_I \rightarrow \mathbb{R}^{n_f}, \, \iota s_I \mapsto \iota f(\iota s_I), \label{eq:sffi} 
\end{align}
which is an sffi. It is important to note in (\ref{eq:sffi}) that the definition domain $\iota \tau s_I$ of $\iota f$ depends on the types $\tau s_I$ of the typed symbols $s_I$ involved in the wff $F(s_I)$ defining the s-function $f\in\mathbb{F}$.
See also Table~\ref{table:typedsymbols}.  
\REVo{ 
Continuing the example, \texttt{symb:i} being of type \texttt{i} (interval), it comes:
\rule{4mm}{0mm}$\iota f_1:$ \scriptsize$[-1,+1] \rightarrow \mathbb{R}^2, \, \sigma \mapsto [0.25+0.5\sigma+0.25\sigma^2; 0.5+0.5\sigma]$\normalsize, \\
\rule{4mm}{0mm}$\iota f_2:$ \scriptsize$[-1,+1] \rightarrow \mathbb{R}, \, \sigma \mapsto 0.25-0.25\sigma^2$\normalsize.
} 

An hybrid example mixing continuous and discrete types is also given with a s-function $f\in\mathbb{F}$ defined as the wff $F(s_I)=$ \texttt{1+x:i+4*y:b*z:s}. Then, $n_f=1$ (scalar output), and the $p_f=3$ involved typed symbols $s_{I_1}=$ \texttt{x:i}, $s_{I_2}=$ \texttt{y:b}, $s_{I_3}=$ \texttt{z:s} are respectively of type \texttt{i} (interval), \texttt{b} (boolean), \texttt{s} (signed). In this example, the symbol names in $F(s_I)$ coincide with those in Table~\ref{table:typedsymbols}. Following (\ref{eq:sffi}), an sffi of $f$ as a mathematical function is:\\
\rule{8mm}{0mm}$\iota f : \inter\times\bool\times\sign \rightarrow \mathbb{R}, \, [x;y;z] \mapsto 1+x+4yz$,\\
and the imset of $\iota \tau s_I$ by $\iota f$ is $[-4,-2]\cup[0,2]\cup[4,6]$ which is a non-convex and non-connected set. This can be generalized with image-sets.

	\subsection{Image-sets} \label{subsec:is}
	
\REVo{ 
Set-based rather than functional (sffi) interpretations of s-functions are considered with the image-sets defined in this paragraph. They provide set-based wraps ensuring the consistency between some intended mathematical meaning and the actually computed sets through an inclusion preserving approach. 
} 
This is obtained by first extending the \emph{imsets} and \emph{inclusion functions} in section~\ref{sec:prelim} to a symbolic context with typed symbols through \emph{image-sets} and \emph{inclusion s-functions}, respectively. Then, a notion of inclusion preserving symbolic reduction operator is introduced.
This gives control to maintain finite representations of prescribed complexity for the underlying approximate polynomial expansions while preserving a set inclusion property ensuring consistency of the computed polynotopic image-sets.   

\begin{defn}[Image-set] \label{defn:imageset}
	The image-set $\imset{f}_{\iota}$ 
	of the s-function $f=F(s_I)\in\mathbb{F}$ is the imset of the domain $\iota \tau s_I$ by a functional interpretation $\iota f$ of $f$: \\ 
	\rule{1cm}{0mm} $\imset{f}_{\iota} = \{ \iota f(\sigma) \,\, | \,\, \sigma \in \iota \tau s_I  \}$ $=\iota f(\iota \tau s_I)$. 
\end{defn}
\begin{rem}
	The domain $\iota \tau s_I$ is a set related to the types of the symbols in $s_I$ (see assumption~\ref{assum:types} and table~\ref{table:typedsymbols}) and 
	$\iota f(\sigma)$ stands for an sffe as in definition~\ref{defn:sfun}. 
\end{rem}
\begin{defn}[Inclusion s-function] \label{defn:inclsfun}
	The s-function $g=G(s_J)$ is an inclusion s-function for $f=F(s_I)$ under functional interpretations $\iota f$ of $f$ and $\iota g$ of $g$ if 
	$\iota g$ is an inclusion function 
	for $\iota f$ defined on the domain $\iota \tau s_I$. 
\end{defn}
\begin{cor} \label{cor:inclsfun}
	From the definition~\ref{defn:inclfun} and its corollary~\ref{cor:inclfuncor2}, 
	${^\subset\iota \tau s_I}$ being a set of subsets of $\iota \tau s_I$ including $\iota \tau s_I$ itself, 
	$\iota f(\Sigma)$ being the imset of $\Sigma$ by $\iota f$, and $\iota g(\Sigma)$ being the image of $\Sigma$ by $\iota g$, it comes:	
	\begin{align} 
		& \forall \Sigma \in {^\subset\iota \tau s_I}, \, \{ \iota f(\sigma) \,\, | \,\, \sigma \in \Sigma  \} =\iota f(\Sigma) \subset \iota g(\Sigma), \label{eq:inclsfun} \\
		& \forall \Sigma \in {^\subset\iota \tau s_I}, \, \iota f(\Sigma) \subset \iota g(\iota \tau s_I).   \label{eq:inclsfun2}
	\end{align}
\end{cor}
\begin{defn}[Reduction] \label{defn:red}
	A reduction is an operator $\downarrow_q$ transforming an s-function $f=F(s_I)$ into an s-function $\bar{f}=\redop{q}f=\bar{F}(s_{\bar{I}})$ such that $\bar{f}$ is an inclusion s-function for $f$ depending on at most $q$ generator symbols: $card(\bar{I})\leq q \in \mathbb{N}$ and $card(.)$ gives the cardinal. \\
\end{defn}
\begin{rem}[Reduction]
	$\bar{I}\cap I\neq\emptyset$ is not mandatory but often useful to limit the inclusion conservatism while controlling the complexity of $\bar{f}$ through its input dimension. 
	In other words, a reduction operator should preserve the more important symbols/dependencies i.e. the ones which significantly contribute to shaping the graph of the mathematical function symbolized by the s-function $f$.
\end{rem}
\begin{exmp}[Inclusion s-function]   \label{exmp:inclsfun}
	In the sample code of Table~\ref{table:samplecode}, the step~8 declares the s-function $f_5$ which is an inclusion s-function for the s-function $f_1$ (resp. $f_3$) declared at step~3 (resp. step~5). Subsequently, $f_6$ (step~9) is also an inclusion s-function for $f_2$ (step~4).
\end{exmp}
\REVo{ 
The example~\ref{exmp:inclsfun} illustrates how (vector) polynomial s-functions like $f_1$ or $f_3$ can be rewritten as (vector) linear ones like $f_5$ while preserving the inclusion property of related set-based interpretations as image-sets. This can even be done automatically by following a general scheme similar to automatic differentiation which makes systematic use of basic operator overloading. 
} 
Pushing further this idea, the graph of non-polynomial functions, possibly including switching ones, can be enclosed within polynotopes, as it will be further addressed in section~\ref{sec:hybridtools}. This approach contributes to significantly enlarge the class of dynamical systems for which reachability and filtering algorithms (like PKF in section~\ref{sec:filtering}) can be readily implemented right from the model equations, just by composing polynotope objects 
$i)$ through intuitive operator/function overloading and 
$ii)$ while preserving the efficiency of uncertainty propagation computations possibly involving continuous, discrete or mixed image-sets.

\section{From symbolic zonotopes and mixed encoding to polynotopes} \label{sec:polynotopes}

The description of specific data structures and algorithms actually implementing methods related to polynotope objects was left opened to a large extent in section~\ref{sec:polynobjwhyhow}, consistently with the encapsulation principle. Such descriptions are the main subject of this section~\ref{sec:polynotopes}.
They are gradually introduced by first dealing with symbolic zonotopes, then mixed encoding, before addressing the more general case of polynotopes.  

	\subsection{Affine s-functions and zonotopes} \label{subsec:linsfun}
	
\begin{defn}[Affine/linear wff] \label{defn:linwff}
	The wff $F(s_I)$ is affine in $s_I$ if it can be written as $c+R s_I$ where the vector $c$ and the matrix $R$ do not depend on the symbolic variables in $s_I$. In particular, it is linear when $c$ is null or can be omitted i.e. $F(s_I)=R s_I$. Shortly,\\
	\rule{1cm}{0mm} Affine wff: $F(s_I) = c+R s_I$.
\end{defn}
\begin{defn}[s-zonotope] \label{defn:szono}
	A symbolic zonotope (s-zonotope) $\zono{f}_{s,\tau}$ is an s-function $f=F(s_I)$ such that the wff $F(s_I)$ is affine in the symbolic variables in $s_I$. 
\end{defn}
\begin{defn}[e-zonotope] \label{defn:zono}
	The e-zonotope related to the s-zonotope $\zono{f}_{s,\tau}$ is the image-set $\imset{f}_{s,\tau,\iota}$ of $f=\zono{f}_{s,\tau}$ under an affine interpretation $\iota f$ of $f$.
	An e-zonotope is thus a set-valued evaluation (semantics) related to a given s-zonotope (syntax). 
\end{defn}
One possible data structure to store a symbolic zonotope is $(c,R,I)$. The related s-function defined by a wff denoted $\zono{c,R,I}_{s,\tau}$ is $f = c + R s_I$, and the related e-zonotope is in (\ref{eq:ezono}): 
\begin{align}
	\imset{c,R,I}_{s,\tau} \,\,\, & = c + R s_I \qquad\qquad\qquad (syntax) \label{eq:szono} \\
	\imset{c,R,I}_{s,\tau,\iota}  & = \{ c + R \sigma \, | \, \sigma \in \iota\tau s_I \} \quad (semantics) \label{eq:ezono}	
\end{align}
The main differences with classical zonotopes defined as $\zono{c,R}=\{c+Rs\,|\,s\in[-1,+1]^n\}$ are twofold:

Firstly, the interplay between syntax and semantics is not caught by the classical definition, whereas it plays a key role in the management of the so-called dependency problem. 
\REVo{   
For example, assuming an entirely continuous case i.e. all the symbols $s_i$ are of type (unit) interval, let us consider the sum $S$ (resp. Minkowski sum $S_\iota$) of the s-zonotopes 
$\imset{0,1,1}_{s,\tau}$ and $\imset{0,-1,1}_{s,\tau}$ (resp. $\imset{0,1,1}_{s,\tau,\iota}$ and $\imset{0,-1,1}_{s,\tau,\iota}$). Then, $S=0+s_1-s_1=0$ (resp. $S_\iota=\inter+\inter=2\inter=[-2,+2]$), and the set-valued interpretation $\iota S=\{0\}$ of $S$ is much less conservative than $S_\iota=[-2,+2]$ while preserving the inclusion property under the considered semantics. Indeed, some uncertainty cancellation has been made possible by formal/symbolic transformations respecting operators syntactic rules, whereas this is no more possible through the Minkowski sum following a set-valued evaluation.
} 
One strategy to improve accuracy thus consists in delaying such set-valued evaluations.
Moreover, whereas $\forall i\in \mathbb{N}, \, \imset{0,1,i}_{s,\tau,\iota}=\zono{0,1}=[-1;+1]$, $\imset{0,1,i}_{s,\tau}=s_i$ is distinct from $\imset{0,1,j}_{s,\tau}=s_j$ as long as $i\neq j$.

From a computational perspective, Matrices with Labelled Columns (MLC) featuring a column-wise sparsity as first introduced in \cite{Combastel2019} lead to efficient implementations of s-zonotope operators such as sum, linear image, interval/box hull, etc. The reader is referred to \cite{Combastel2019} for a detailed description, especially in the sections 4 ``Matrices with Labeled Columns (MLC)'' and 6 ``Symbolic zonotopes''. Notice that the definition of symbolic zonotopes in \cite{Combastel2019} only considers one type of symbols interpreted as random variables with support in the unit interval $[-1,+1]$. This outlines how the approach described in section~\ref{sec:framework} can also be used in a stochastic paradigm: Indeed, it suffices to consider other types of symbols interpreted as random variables (which are themselves functions, so emphasizing the relevance of cross-connections with functional paradigms). In order to give a flavor about MLC, an informal definition and a sum example are provided.  
\REVo{ 
An MLC $M^{|I}$ is a pair $(M,I^T)$ where $M$ is an $n\times p$ matrix and $I\in \mathbb{N}^p$ is a vector such that each scalar $I_j$ for $j=1,\ldots,p$, uniquely identifies the $j$th column $M_{:,j}$ of $M$. Shortly, the column of $M^{|I}$ labeled as $I_j$ refers to $M_{:,j}$. An example illustrating the sum of two MLC, $M^{|I}+N^{|J}=P^{|K}$ is:
\begin{small}
	\begin{equation} \label{eq:MLCsumexmpl:main}
		\left[
		\begin{array}{ccc}
		2 & 1 & 5 \\ 
		\hline
		1 & 0 & 1 \\ 
		0 & 1 & 1
		\end{array}
		\right]
		+
		\left[
		\begin{array}{cccc}
		3 & 5 & 2 & 8 \\ 
		\hline
		2 & 0 & 4 & 6 \\ 
		0 & 3 & 5 & 7 \\
		\end{array}
		\right]	
		=
		\left[
		\begin{array}{ccccc}
		1 & 2 & 5 & 3 & 8 \\ 
		\hline
		0 & 5 & 1 & 2 & 6  \\ 
		1 & 5 & 4 & 0 & 7
		\end{array}
		\right].			
	\end{equation}
	\begin{equation} \label{eq:MLCsumprop:main}
		\begin{bmatrix}
		0 & 5 & 1 & 2 & 6 \\ 
		1 & 5 & 4 & 0 & 7			
		\end{bmatrix}
		\begin{bmatrix}
		\rand{s}_1 \\ \rand{s}_2 \\ \rand{s}_5 \\ \rand{s}_3 \\ \rand{s}_8				
		\end{bmatrix}
		=
		\begin{bmatrix}
		1 & 0 & 1 \\ 
		0 & 1 & 1			
		\end{bmatrix}
		\begin{bmatrix}
		\rand{s}_2 \\ \rand{s}_1 \\ \rand{s}_5			
		\end{bmatrix}				
		+
		\begin{bmatrix}
		2 & 0 & 4 & 6 \\ 
		0 & 3 & 5 & 7			
		\end{bmatrix}
		\begin{bmatrix}
		\rand{s}_3 \\ \rand{s}_5 \\ \rand{s}_2 \\ \rand{s}_8			
		\end{bmatrix}
	\end{equation}
\end{small}%
As shown in (\ref{eq:MLCsumexmpl:main}), an equal number of columns/generators of the operands is not mandatory, and the labels are reported on the first lines (e.g. $I^T=[2, 1, 5]$). (\ref{eq:MLCsumprop:main}) illustrates the column-wise (not row-wise) sparsity granted by MLC operators, and that the sum of two MLC gives the exact sum of two centered s-zonotopes since $\zono{0,P,K}_{s,\tau} = \zono{0,M,I}_{s,\tau} + \zono{0,N,J}_{s,\tau}$. Indeed, $M^{|I}+N^{|J}=P^{|K} \Rightarrow Ms_I+Ns_J=Ps_K$. $K\subset I \cup J$ results from merging the unique identifiers in $I$ and $J$ while removing those possibly related to null generators. 
The vertical concatenation $[M^{|I}; N^{|J}]=[M^{|I}; 0]+[0; N^{|J}]$ also illustrates close links between sum and concatenation of MLC.
} 

\begin{figure}
	\begin{center}
		\includegraphics[width=5cm]{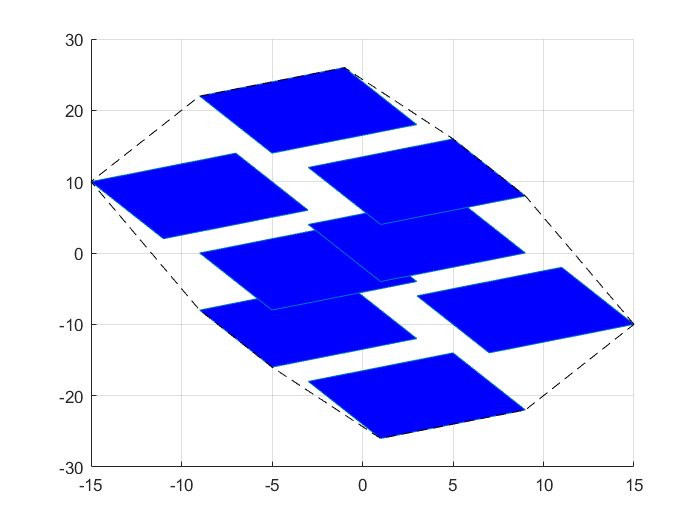}
	\end{center}
	\caption{Example of mixed e-zonotope $\imset{0,R,I}_{s,\tau,\iota}$.}
	\label{fig:mezono}
\end{figure}

Secondly, another main difference with classical zonotopes is the introduction of symbol typing. This makes it possible to combine several kinds of interpretations. Following the assumption~\ref{assum:types}, this paper mainly focuses on mixed, continuous and discrete values in a set-membership paradigm, though the approach described in section~\ref{sec:framework} is more general. 
For example, let us consider five symbols $s_i$, $i=1,\ldots,5$, such that $s_1, s_2, s_3$ are of type $\mathtt{s}$ (signed) i.e. $s_1, s_2, s_3$ take their values in the discrete set $\{-1,+1\}$, and $s_4, s_5$ are of type $\mathtt{i}$ (interval) i.e. $s_4, s_5$ take their values in the unit interval $[-1,+1]\subset \mathbb{R}$. Let $I_1=[1,2,3]$ and $I_2=[4,5]$ be vectors of unique identifiers gathering symbols according to their type: Following the definition~\ref{defn:mcd}, $s_{I_1}$ is discrete, $s_{I_2}$ is continuous, and $s_I$ with $I=[I_1; I_2]$ is mixed. 
Let $R_1=[3, 3, -3; 6, -5, 9]$, $R_2=[4, 2; 2, -4]$ and $R=[R_1, R_2]$.
Then,
the (discrete) e-zonotope $\imset{0,R_1,I_1}_{s,\tau,\iota}$ is a set of eight\footnote{It could be less if some image points are identical.} points which are the linear images by $R_1$ of the vertices of a 3D unit hypercube since $\iota s_{I_1} \in \iota\tau s_{I_1}=\{-1,+1\}^3$.
The (continuous) e-zonotope $\imset{0,R_2,I_2}_{s,\tau,\iota}$ is the classical zonotope $\imset{0,R_2}$ i.e. the linear image by $R_2$ of a 2D unit hypercube since $\iota s_{I_2} \in \iota\tau s_{I_2}=[-1,+1]^2$.
More interestingly, the (mixed) e-zonotope  $\imset{0,R,I}_{s,\tau,\iota}$ is the linear image by $R=[R_1, R_2]$ of $\iota\tau s_I = [\{-1,+1\}^3;[-1;+1]^2]$. It is also the Minkowski sum of the (discrete) e-zonotope $\imset{0,R_1,I_1}_{s,\tau,\iota}$ and the (continuous) e-zonotope $\imset{0,R_2,I_2}_{s,\tau,\iota}$. The resulting set is neither convex nor connected, as shown in Fig.~\ref{fig:mezono}, where the dashed line is the border of the classical (continuous) zonotope $\zono{0,R}$.
This example illustrates that mixed zonotopes can provide a very compact representation (Fig.\ref{fig:mezono}: $R\in\mathbb{R}^{2\times5}$ and 5 bits encoding the symbol types) for the union of a same\footnote{Relaxing this constraint is among the motivations to the polynomial extension in \S\ref{subsec:polsfun}.\label{ftn:motivpoly}} (continuous) zonotopic shape centered on each point of a discrete set possibly containing a high number of configurations (Fig.~\ref{fig:mezono}: $2^3=8=$ cardinal of $\{-1,+1\}^3$). 

	\subsection{Mixed encoding} \label{subsec:mxenc} 

The notion of \emph{mixed encoding} is introduced in the same spirit as the example illustrated in Fig.~\ref{fig:mezono}. It also provides an approach for a hierarchical modeling of dependencies making it possible to tune the granularity level of the description. Notation: Let $\inter_i$ (resp. $\sign_i$, $\bool_i$) denote the symbol $s_i$ provided it is of type interval (resp. signed, boolean) i.e. $\tau s_i=\mathtt{i}$ (resp. $\mathtt{s}$, $\mathtt{b}$). Under the assumption~\ref{assum:types}, a compact notation for typed symbols is so obtained, each being uniquely identified by $i$.
Also, let $\rho(0)=1$, $\rho(n+1)=\frac12[1, \rho(n)]$ for $n\in\mathbb{N}$. Then, $\forall n$, $\rho(n)=[(\frac12)^1, \ldots, (\frac12)^n, (\frac12)^n] \in \mathbb{R}^{1\times(n+1)}$. By induction, the row sum of $\rho(n)$ is $1$. 
\begin{defn}[Mixed encoding of basic intervals] \label{defn:mxenc}
	The s-zonotope $Z_s^n(I)$ is an $n$-level signed-interval mixed encoding of the unit interval $[-1,+1]$ if \\ 
	\rule{2mm}{0mm} $Z_s^n(I) = \zono{0,\rho(n),I}_{s,\tau} = (\sum_{j=1}^n (\frac12)^j \sign_{I_j}) + (\frac12)^n \inter_{I_{n+1}}$. \\
	The s-zonotope $Z_b^n(I)$ is an $n$-level boolean-interval mixed encoding of the interval $[0,1]$ if \\ 
	\rule{2mm}{0mm} $Z_b^n(I) = \zono{0,\rho(n),I}_{s,\tau} = (\sum_{j=1}^n (\frac12)^j \bool_{I_j}) + (\frac12)^n \inter_{I_{n+1}}$. \\
\end{defn}
\begin{cor}[Mixed encoding of intervals] \label{cor:mxenc_int}
	Let $Z_s^n(I)$ be a mixed encoding of $[-1,+1]$. Then, $c + r Z_s^n(I)$ is a mixed encoding of $c\pm r = [c-r, c+r]$. \\
	Let $Z_b^n(I)$ be a mixed encoding of $[0,1]$. Then, $a + (b-a) Z_b^n(I)$ is a mixed encoding of $[a, b]$.
\end{cor}
\begin{cor}[Related e-zonotopes] \label{cor:mxenc_ezono}
	Following the definition~\ref{defn:mxenc}, the e-zonotope related to the s-zonotope $Z_s^n(I)$ (resp. $Z_b^n(I)$) is $[-1,+1]$ (resp. $[0,1]$). Conversely, there is no unique mixed encoding for a given interval. \\ 
	The e-zonotope $(c + r Z_s^n(I))_\iota$ related to the s-zonotope $c + r Z_s^n(I)$ satisfies $(c + r Z_s^n(I))_\iota \subseteq (c\pm r)$ (interval hull). The equality holds in the scalar case or for $r=0$.
\end{cor}
The surjective nature of mixed encoding gives freedom degrees to model dependencies in a hierarchical way. The discrete parts feature close analogies with the usual binary encoding of integers. Moreover, the coverage of continuous domains is achieved through remainder terms. The width of the set-valued interpretation of these terms is related to the granularity of the mixed-encoding. It can be refined or reduced by adapting the level value $n$. \\
Since affine s-functions and zonotopes essentially provide operators managing affine dependencies only, and since this yields some restrictions on the possible uses of mixed encoding (among others: see, e.g., footnote~\ref{ftn:motivpoly}), an extension to polynomial dependencies is considered.    

	\subsection{Polynomial s-functions and polynotopes} \label{subsec:polsfun}

Let $(\boxplus,\boxtimes)$ denote a generic matrix product: 
$M (\boxplus,\boxtimes) N = \boxplus_{j=1}^{p}(M_{ij} \boxtimes N_{jk})$, where $p$ both refers to the number of columns of $M$ and the number of rows of $N$. For example, $MN = M(+,.)N$ is the classical matrix product. 
$+$, $.$, $\verb|^|$ respectively denote sum, product, power.
\begin{defn}[Monomial matrix notation]
	The monomial matrix $\theta^E$ is $(\theta^T (.,\verb|^|) E)^T$, where $\theta$ (resp. $E$) is a so-called variable matrix (resp. exponent matrix) of dimension compatible with the generic matrix product $(.,\verb|^|)$. The operator $\,^T$ refers to transposition. 
\end{defn}
Examples: Taking $\theta=[s_1; s_2]$ and $E=[1, 0, 2; 0, 3, 4]$ yields $\theta^E=[s_1; s_2^3; s_1^2 s_2^4]$. $\theta^\identity=\theta$ with $\identity=$ identity.\\
\begin{defn}[Polynomial wff] \label{defn:linwff}
	The wff $F(s_I)$ is polynomial in $s_I$ if it can be written as $c+R s_I^E$ where the vector $c$ and the matrices $R$ and $E$ do not depend on the symbolic variables in $s_I$. Shortly,\\
	\rule{1cm}{0mm} Polynomial wff: $F(s_I) = c+R s_I^E$.
\end{defn}
Then, $c$, $R$, $s_I$, $E$, $s_I^E$ are respectively the so-called constant vector, coefficient/generator matrix, symbol(ic variable) vector, exponent matrix, monomial vector.
\begin{defn}[s-polynotope] \label{defn:spoly}
	A symbolic polynotope (s-polynotope) $\zono{f}_{s,\tau}$ is an s-function $f=F(s_I)$ such that the wff $F(s_I)$ is polynomial in the symbolic variables in $s_I$. 
\end{defn}
\begin{defn}[e-polynotope] \label{defn:poly}
	The e-polynotope related to the s-polynotope $\zono{f}_{s,\tau}$ is the image-set $\imset{f}_{s,\tau,\iota}$ of $f=\zono{f}_{s,\tau}$ under a polynomial interpretation $\iota f$ of $f$.
	An e-polynotope is thus a set-valued evaluation (semantics) related to a given s-polynotope (syntax). 
\end{defn}
The name polynotope introduced in this work originates from a contraction of polynomial and zonotope. Following~\cite{Kochdumper2019a}, 
it is also willingly close to polytope. So, \emph{polynotope} gathers, at least partially, the Greek roots of polynomial (from \emph{polus}:numerous and \emph{nomos}:division) and polytope (from \emph{polus} and \emph{topos}:location). \\
Under the assumption~\ref{assum:types}, polynotopes are to sparse polynomial zonotopes what mixed zonotopes are to zonotopes. They can also be viewed as constrained polynomial zonotopes with polynomial constraints managed through a possibly extensible set of symbol types.

\begin{figure}
	\begin{center}
		\includegraphics[width=5cm]{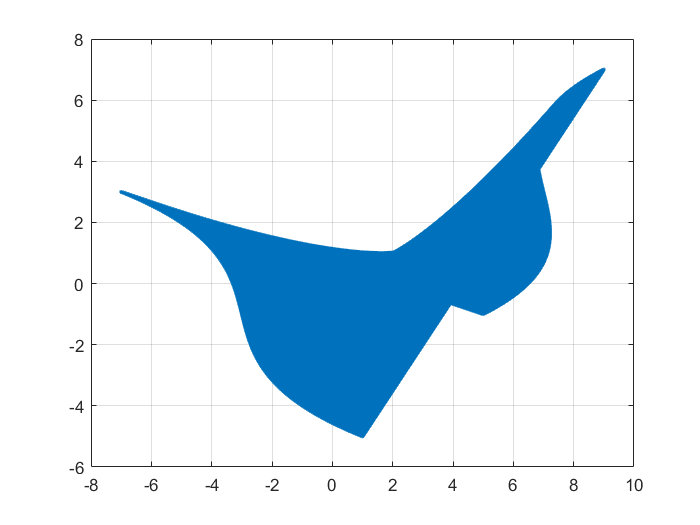}
	\end{center}
	\caption{Example of continuous e-polynotope $\imset{c,R,I,E}_{s,\tau,\iota}$.}
	\label{fig:cepoly}
\end{figure}

One possible data structure to store a symbolic polynotope is $(c,R,I,E)$. The related s-function defined by a wff denoted  $\zono{c,R,I,E}_{s,\tau}$ is $f=(I, c + R s_I^E)$, and the related e-polynotope is in (\ref{eq:epoly}): 
\begin{align}
	\hspace{-3mm} \imset{c,R,I,E}_{s,\tau} \,\,\, & = c + R s_I^E \qquad\qquad\quad\,\,\,\, (syntax) \label{eq:spoly} \\
	\hspace{-3mm} \imset{c,R,I,E}_{s,\tau,\iota}  & = \{ c + R \sigma^E \, | \, \sigma \in \iota\tau s_I \} \, (semantics) \label{eq:epoly}	
\end{align}
Polynotopes as in (\ref{eq:spoly})$-$(\ref{eq:epoly}) generalize zonotopes as in (\ref{eq:szono})$-$(\ref{eq:ezono}). Indeed, zonotopes are obtained for $E=\identity$ (identity matrix) which is highly sparse and can thus be stored very efficiently. Using a sparse $E$ with integer entries in $\mathbb{N}$ leads to a data structure similar to sparse polynomial zonotopes (spz) in \cite{Kochdumper2019}, where no typing of symbols is considered (continuous case only). The sparsity of $E$ extends the column-wise sparsity of MLC to polynomial (rather than affine) dependencies with a compact description of monomials featuring (almost) no restriction on the highest degree. The example (\ref{eq:spolydataexmp})$-$(\ref{eq:spolysfunexmp}) shows how the s-polynotope related to (\ref{eq:spolysfunexmp}) can be compactly encoded using $(c,R,I,E)$ as in (\ref{eq:spolydataexmp}). See also Fig.~\ref{fig:cepoly}.
\begin{align}
	& \left[\begin{array}{c|c} I & E \\ \hline c & R \end{array}\right]
	= 
    \left[\begin{array}{c|ccc} 
    	1 & 1 & 0 & 1 \\ 
    	8 & 0 & 3 & 1 \\
    	\hline 
    	2 & 5 & 3 & -1 \\
    	1 & 2 & 0 & 4
    \end{array}\right], \label{eq:spolydataexmp} \\
    & \left[\begin{array}{l}
    	s_1 \\ 
    	s_8 
    \end{array}\right]
    \mapsto
    \left[\begin{array}{l}
    	2 + 5 s_1 + 3 s_8^3 - s_1 s_8 \\ 
    	1 + 2 s_1 + 4 s_1 s_8 
    \end{array}\right]. \label{eq:spolysfunexmp}
\end{align}
The implementation of continuous polynotopes operations used in this work is close to the one described in \cite{Kochdumper2019} for spz. In particular, each time monomial redundancies might occur, they are removed by summing the related generators: all the columns of $E$ remain distinct. The main differences are: \\
1) The case of independent generators is not treated separately i.e. all the generators are possibly dependent (provided they share some common symbol), \\
2) The implementation of a symbolic addition is considered and optimized by taking into account the fact that monomials/generators involving at least one own variable from an operand can be simply copied in the result since no similar monomial exists in the other operand, \\
3) A vertical concatenation extends the one of MLC, \\ 
4) An element-wise product is used as a special case of quadratic map and the reduction extends the one in \cite{Combastel2019}.

Our implementation of polynotopes also supports discrete and mixed operations through symbol typing as described in section~\ref{sec:framework} and assumption~\ref{assum:types}. Compared to a strictly continuous case as in \cite{Kochdumper2019}, the main difference is the introduction of rewriting rules taking the specific nature of signed and boolean symbols into account. 
Related substitutions ($\rightarrow$) are implemented very efficiently using the $(c,R,I,E)$ attributes with sparse $E$, e.g.
\begin{align}
	& (\sign_i)^n \rightarrow (\sign_i)^{mod(n,2)}, \label{eq:rrspow} \\ 
	& (\bool_i)^n \rightarrow (\bool_i)^{max(n,1)}. \label{eq:rrbpow} 
\end{align}
\begin{defn}[Rewriting rules and inclusion] \label{defn:rrincl}
	A rewriting rule is inclusion preserving if: \\ 
	\rule{10mm}{0mm} $(\imset{f}_{s,\tau} \rightarrow \imset{g}_{s,\tau}) \Rightarrow (\imset{f}_{s,\tau,\iota} \subseteq \imset{g}_{s,\tau,\iota})$. \\
	It is inclusion neutral if: \\
	\rule{10mm}{0mm} $(\imset{f}_{s,\tau} \rightarrow \imset{g}_{s,\tau}) \Rightarrow (\imset{f}_{s,\tau,\iota} = \imset{g}_{s,\tau,\iota})$.
\end{defn}
\begin{prop} \label{prop:rrprod}
	The rewriting rules in (\ref{eq:rrspow}) and (\ref{eq:rrbpow}) are inclusion neutral under the assumption~\ref{assum:types}. 
\end{prop}
Notice the syntactical (resp. semantic) nature of the premises (resp. conclusions) of the implications ($\Rightarrow$) in the definition~\ref{defn:rrincl}.
To give insight into the proposition~\ref{prop:rrprod}, let $x$ be a possible value of any signed symbol: $x\in\{-1,+1\}\subset\mathbb{R}$. Thus, $(x+1)(x-1)=0$ i.e. $x^2=1$ as polynomial constraint. By induction, $x^n=x$ for odd $n$, $x^n=1$ for even $n$, that is $x^n = x^{mod(n,2)}$ which shows the inclusion neutrality of (\ref{eq:rrspow}). Similarly, let $x$ be a possible value of any boolean symbol: $x\in\{0,1\}\subset\mathbb{R}$. Thus, $(x-0)(x-1)=0$ i.e. $x^2=x$. By induction, $x^n=x$ if $n>1$, $x^n=1$ if $n=0$, that is $x^n = x^{max(n,1)}$ which shows the inclusion neutrality of (\ref{eq:rrbpow}). \\
The rewriting rules (\ref{eq:rrspow})$-$(\ref{eq:rrbpow}) apply for operations modifying the monomial degrees like product; the number of distinct monomials induced by discrete operations is drastically reduced compared to continuous ones, since the exponent in the right term of (\ref{eq:rrspow})$-$(\ref{eq:rrbpow}) is either $0$ or $1$ instead of any $n\in\mathbb{N}$. Thanks to inclusion neutrality, such simplifications of formal expressions induce no conservatism in the related set-valued interpretations.
Other rewriting rules are only inclusion preserving:
\begin{prop} 
	The rewriting rules in (\ref{eq:rrb2s}), (\ref{eq:rrs2i}) and (\ref{eq:rri2i}) are inclusion preserving under the assumption~\ref{assum:types}. 
	\begin{align} 	
		& (\bool_i) \rightarrow 1/2 + (\sign_j)/2,     \label{eq:rrb2s} \\
		& (\sign_i) \rightarrow (\inter_j),    	       \label{eq:rrs2i} \\
		& (\inter_i)^2 \rightarrow 1/2 + (\inter_j)/2 , \quad (\inter_i)(\inter_j) \rightarrow (\inter_k).  \label{eq:rri2i}
	\end{align}
\end{prop}
(\ref{eq:rrb2s})$-$(\ref{eq:rri2i}) apply before computing the zonotope/interval enclosure of a (possibly mixed) polynotope. Notice that (\ref{eq:rrb2s}) is  inclusion neutral if applied globally i.e. without generating new symbol multi-occurrences. (\ref{eq:rrs2i}) is the formal/syntactical counterpart of $\{-1,+1\}\subset[-1,+1]$ which can be viewed as a prototype of the most basic inclusion of two discrete modes/configurations ($-1$ and $+1$) into a single continuous domain (the unit interval). \\
By using appropriate reductions of formal expressions based on inclusion preserving rewriting rules, mixed polynotopes provide a highly versatile, scalable and computationally efficient approach to combine and enclose possibly non convex and non connected sets under dependency constraints. 
Hence, they look appealing to deal with verification and synthesis of Cyber-Physical Systems (CPS) whose modeling often relies on mixed/hybrid dynamics. They also exemplify the generality of the approach of image sets with typed symbols described in section~\ref{sec:framework}.

\section{Modeling tools for nonlinear hybrid systems} \label{sec:hybridtools}

	\subsection{Discrete: Signed and Boolean logic functions} \label{subsec:discSBfun}

This paragraph shows how signed (resp. boolean) symbolic variables can be used in a polynomial framework, like the one of polynotopes under the assumption~\ref{assum:types}, to express any propositional logic formula where symbolic variables are interpreted on a bi-valued real domain: 
$\sign=\{-1,+1\}\subset\mathbb{R}$ (resp. $\bool=\{0,1\}\subset\mathbb{R}$). This gives a natural interface between continuous variables (defined on a (real) domain with infinite cardinal) and discrete ones (defined on a (real) domain with finite cardinal). 

\begin{prop}[Multi-affine decomposition] \label{prop:multiaffdecomp}		
	Let $f:\mathbb{R}^p \rightarrow \mathbb{R}^n$ be any function between finite dimensional real domains. Let $x\in\mathbb{R}$ and $z\in\mathbb{R}^{p-1}$ so that $(x,z)\in\mathbb{R}^p$ refers to any input vector of $f$ where a scalar input $x$ is distinguished from the others. Let introduce four partial functions of $f$ defined as: \\
	\begin{tabular}{@{\rule{6mm}{0mm}}l@{$\,:\,$}l}
		$f^A_x(z)=\frac{f(+1,z)+f(-1,z)}2$ & Average of $f$ wrt $x$, \\
		$f^H_x(z)=\frac{f(+1,z)-f(-1,z)}2$ & Half-gap of $f$ wrt $x$, \\
		$f^G_x(z)=f(0,z)$        & Ground of $f$ wrt $x$, \\
		$f^U_x(z)=f(1,z)-f(0,z)$ & Unit-gap of $f$ wrt $x$, 
	\end{tabular}\\	
	Then, an affine decomposition of $f$ wrt $x$ under a signed (resp. boolean) $x$ is respectively given by (\ref{eq:signdecomp}) and (\ref{eq:booldecomp}). Moreover, if all the scalar entries of $z$ are signed (resp. boolean), a recursive application of (\ref{eq:signdecomp}) (resp. (\ref{eq:booldecomp})) results in a (polynomial) multi-affine decomposition of $f$.  
	\begin{align}
		& x\in\{-1,+1\} \Rightarrow f(x,z)=f^A_x(z)+x f^H_x(z), \label{eq:signdecomp} \\
		& x\in\{\color{white}-\color{black}0,\color{white}+\color{black}1\} \Rightarrow f(x,z)=f^G_x(z)+xf^U_x(z). \label{eq:booldecomp}
	\end{align}
\end{prop}
\textbf{Proof.} (\ref{eq:signdecomp}) comes from $f(+1,z)=f^A_x(z)+f^H_x(z)$ and $f(-1,z)=f^A_x(z)-f^H_x(z)$. Similarly, (\ref{eq:booldecomp}) comes from $f(0,z)=f^G_x(z)$ and $f(1,z)=f^G_x(z)+f^U_x(z)$. $\hspace{\stretch{1}}\square$ \\
The multi-affine decomposition of basic propositional logic operators is reported in Table~\ref{table:logfun} both in the signed and boolean cases. 
	At least three noticeable facts emerge from Table~\ref{table:logfun}: \\
	$a)$ The equivalence \emph{eqv} in the signed case features the same multi-affine decomposition as the logical \emph{and} in the boolean case, and both reduce to a simple product. \\
	$b)$ The multi-affine decompositions with signed operands look more ``balanced'' in terms of involved monomials, compared to the boolean case. This is visible right from the basic affine decompositions in (\ref{eq:signdecomp}) and (\ref{eq:booldecomp}). Indeed, the average of both alternatives (resp. the $0$ alternative) serve as reference to express the impact of a switching controlled by $x$ in the signed (resp. boolean) case. \\
	$c)$ Interpreting in $\mathbb{R}$ the polynomial expression of a multi-affine decomposition yields some interpolation between discrete configurations initially expressed in a (bi-valued) propositional logic framework.   
\begin{table}	
	\begin{center}
		\caption{Signed and Boolean logic functions related to basic operators expressed in the ring of multivariate 
		polynomials $\mathbb{R}[s_I]$ with coefficients in the real field $(\mathbb{R},+,.)$.} \label{table:logfun}
		\begin{tabular}{lcll}
			& $(a,b)$ & $\in\{-1,+1\}^2$ & $\in\{0,1\}^2$ \\ 
			\hline
			Op.  & Symb. & Signed & Boolean \\
			\hline
			not  & $\neg$ & $-a$ & $1-a$ \\
			and  & $\wedge$ & $\frac{-1+a+b+ab}2$ & $ab$\\
			or   & $\vee$ & $\frac{+1+a+b-ab}2$ & $a+b-ab$ \\
			nand & $\uparrow$, $\barwedge$ & $\frac{+1-a-b-ab}2$ & $1-ab$ \\
			nor  & $\downarrow$, $\veebar$ & $\frac{-1-a-b+ab}2$ & $1-a-b+ab$ \\
			imp  & $\Rightarrow$, $\le$ & $\frac{+1-a+b+ab}2$ & $1-a+ab$ \\
			eqv  & $\Leftrightarrow$, $=$ 
			     & $\color{white}+\color{black}ab$ & $1-a-b+2ab$ \\
			xor  & $\nLeftrightarrow$, $\neq$ 
				 & $-ab$ & $\color{white}0+\color{black}a+b-2ab$\\		
			\hline
			pow  & $a^n, n\in\mathbb{N}$ & $a^{mod(n,2)}$ & $a^{max(n,1)}$ \\
			true  & $\top$ & $+1$ & $1$ \\
			false & $\bot$ & $-1$ & $0$ \\ 
			\hline
		\end{tabular}
	\end{center}
\end{table}
\begin{prop}[Logical ordering] \label{prop:orderops}
	Let $(a,b)$ be a pair of signed (resp. boolean) symbolic variables. Defining the operator $>$ such that $(a > b) = \neg (a \le b)$ holds true with operators as in table~\ref{table:logfun}, then $(a < b) = (b > a)$ and $(a \le b) = ((a < b) \vee (a = b))$ also hold true. More generally, the operators $\le$ (i.e. implication\footnote{Notice that contraposition writes as $(a \le b) = (\neg b \le \neg a)$.}), $\ge$, $>$, $<$ follow similar rules as classical order relation operators over reals when signed (resp. boolean) symbols are interpreted with values in $\{-1,+1\}\subset\mathbb{R}$ (resp. $\{0,1\}\subset\mathbb{R}$). 
\end{prop}
\begin{thm}[Functional completeness] \label{thm:completeness} 
	Under the assumption~\ref{assum:types}, let $I\subset\mathbb{N}$ be a finite set of unique symbol identifiers with at least $p$ elements of type signed (resp. boolean). s-polynotopes based on wff interpreted as multivariate polynomials $\mathbb{R}[s_I]$ with coefficients in the real field $(\mathbb{R},+,.)$ can describe any function $f:\sign^p \rightarrow \sign$ (resp. $f:\bool^p \rightarrow \bool$), where $\sign=\{-1,+1\}\subset\mathbb{R}$ (resp. $\bool=\{0,1\}\subset\mathbb{R}$). 
\end{thm}
	\textbf{Proof.}
	Theorem~\ref{thm:completeness} follows from the functional completeness of the nand (or nor) logical operator and the fact that the composition of polynomials in $\mathbb{R}[s_I]$ result in polynomials in $\mathbb{R}[s_I]$. 
	Indeed, the nand operator is defined as a polynomial function with signed (resp. boolean) operands and codomain in Table~\ref{table:logfun}. Thus, the composition of any number of such nand operations on signed (resp. boolean) symbolic variables evaluated in $\sign$ (resp. $\bool$) result in a polynomial s-function i.e. a s-polynotope according to the definition~\ref{defn:spoly}.    
\begin{cor}
	Given any pair $(a,b)\in\mathbb{R}^2$ satisfying $a<b$, all the finite dimensional Boolean functions and operations can be plunged in $\{a,b\}^p \subset \mathbb{R}^p$ for some $p\in\mathbb{N}$ after suitable re-scaling compared to the usual Boolean case i.e. $(a,b)=(0,1)$. This is exemplified with the so-called signed case i.e. $(a,b)=(-1,+1)$ in table~\ref{table:logfun}, where the direct and inverse affine re-scaling functions are $r:\sign \rightarrow \bool, x \mapsto \frac{1+x}2$ and $r^{-1}: \bool \rightarrow \sign, x \mapsto 2x-1$.  
\end{cor}

	\subsection{Continuous: Nonlinear functions} \label{subsec:contNLfun}

Whereas the imset (see definition~\ref{defn:imset}) of a polynotope (resp. zonotope) by a polynomial (resp. affine) function is still a polynotope (resp. zonotope), the imset by non-polynomial (resp. non-linear) functions is not. This paragraph proposes a method to obtain guaranteed inclusions of non-polynomial (resp. non-linear) functions while maintaining dependency links between inputs and outputs. Indeed, breaking such links (e.g. by a naive use of interval arithmetic) is often the source of over-approximations and/or wrapping effect.

For the sake of simplified notations, the distinction between s-functions (syntax) and their usual interpretation as a mathematical function (semantic) is not systematic in the following. Example: Let $f(x)=e^x$. Then, $f$ refers either to the s-function built from the wff $e^x$, or to $\iota_m f:x \mapsto e^x$. \\
The notations used for intervals are as follows: $x\in[x]=[\binf{x}, \bsup{x}]=\bmid{x}\pm\brad{x}=[\bmid{x}-\brad{x}, \bmid{x}+\brad{x}]$ where $\binf{x}$, $\bsup{x}$, $\bmid{x}$, $\brad{x}$ respectively denote the lower bound, upper bound, center (or middle), radius of the interval $[x]$ containing $x$. Then, $\bmid{x}=\frac{\bsup{x}+\binf{x}}2$ 
and $\brad{x}=\frac{\bsup{x}-\binf{x}}2$. 
Recall: $\inter=[-1,+1]$.
\begin{lem}[Unit range mapping] \label{lem:unitmap}
	Let $[x]=[\binf{x}, \bsup{x}]=\bmid{x}\pm\brad{x}\subset\mathbb{R}$.
	Let $\mu: [x] \rightarrow \inter, \, x \mapsto \delta=\frac{x-\bmid{x}}{\brad{x}}$ if $\brad{x}\neq0$, $\delta=0$ otherwise.
	$\mu^{-1}(\delta)=\bmid{x}+\brad{x}\delta$.
	Unless $\brad{x}=0$ (degenerate point case), the unit range mapping $\mu$ (or $\mu_{[x]}$) of $[x]$ is linear and bijective: It maps the interval range $[x]$ of $x$ to the unit interval $\inter$ containing any $\delta=\mu(x)$, $x\in[x]$.
\end{lem}
Given an interval $[x]\subset\mathbb{R}$, let $f: [x] \rightarrow  \mathbb{R}, \, x \mapsto y=f(x)$ be a function that does not satisfy a property $\pi$ (i.e. $\neg\pi(f)$ is true) required for a given class of image-sets to be closed under the (element-wise) application of $f$ to the underlying s-functions. For example: $\pi$ being the property of being linear (resp. polynomial), the class of e-zonotopes (resp. e-polytopes) are closed under the application of linear (resp. polynomial) functions to the underlying s-zonotopes (resp. s-polytopes). $\neg\pi(f)$ then means that $f$ is non-linear (resp. non-polynomial) for zonotopes (resp. polynotopes). The considered structural property $\pi$ is assumed to be preserved through function composition. 
\begin{lem}[Generic inclusion method] \label{lem:geninclmeth}
	Given an interval $[x]\subset\mathbb{R}$ and $f: [x] \rightarrow \mathbb{R}, \, x \mapsto y=f(x)$ with $\neg\pi(f)$. Let $g:\inter^2 \rightarrow \mathbb{R}, \, (\delta,\epsilon) \mapsto g(\delta,\epsilon)$ be a function satisfying $\forall x\in[x]$, $\exists \epsilon\in\inter$, $f(x)=g(\mu(x),\epsilon)$, where $\mu$ is the unit range mapping of $[x]$.
	Then, $\tilde{f}(.)=g(\mu(.),\inter)$ is an inclusion function for $f(.)$.
	Also, $\pi(g) \wedge \pi(\mu) \Rightarrow \pi(\tilde{f})$.
\end{lem}
According to Lemma~\ref{lem:geninclmeth}, enclosing a non-linear (resp. non-polynomial) function $f$ in a linear (resp. polynomial) framework can be achieved by finding an adequate linear (resp. polynomial) function $g$. In order to exemplify the generic inclusion method, a more focused approach is proposed for increasing/decreasing and convex/concave functions $f$ on some interval $[x]$.
Notation: $[\partial_x f](.) = \left.\frac{\partial f(x)}{\partial x}\right|_{x=.}$.
\begin{thm}[An inclusion method] \label{thm:lininclmeth}
	Let $f: [x] \rightarrow \mathbb{R}, \, x \mapsto y=f(x)$ be a class $\mathcal{C}^1$ 
	convex or concave function on a given interval $[x]=\bmid{x}\pm\brad{x}=[\binf{x}, \bsup{x}] \subset\mathbb{R}$ with $\brad{x}>0$.
	Let $\,$ $\binf{y}=f(\binf{x})$, $\bsup{y}=f(\bsup{x})$, $\bmid{y}=(\bsup{y}+\binf{y})/2$, $\brad{y}=(\bsup{y}-\binf{y})/2$. 
	Let  $\delta=\mu(x)$ where $\mu$ is the unit range mapping of $[x]$ (so, $x\in[x] \Leftrightarrow \delta\in\inter$).
	Let $r(\delta)=f(\bmid{x} + \brad{x} \delta) - (\bmid{y} + \brad{y} \delta)$.
	Let $\delta^*$ be the solution of $[\partial_x f](\bmid{x} + \brad{x} \delta) = \brad{y}/\brad{x}$. Then, $g(\delta,\epsilon)=g_0+g_1\delta+g_2\epsilon$ 
	with $g_0 = \bmid{y}+\frac12 r(\delta^*)$, $g_1=\brad{y}$, $g_2=\frac12 |r(\delta^*)|$ satisfies $\forall x\in[x]$, $\exists \epsilon\in\inter$,
	$f(x)=g(\mu(x),\epsilon)$.  
	$\tilde{f}(.)=g(\mu(.),\inter)$ is an inclusion function for $f(.)$ on $[x]$.
\end{thm}
	\textbf{Proof.}
	The regularity of $f$ on $[x]$ ensures that $[\partial_\delta r](\delta^*) = 0$ i.e. $[\partial_x f](\bmid{x} + \brad{x} \delta^*) = \brad{y}/\brad{x}$ has a unique solution. 
	Since $r(-1)=r(+1)=0$, if $f$ is convex (resp. concave) on $[x]$, then $r(\delta)\in[r(\delta^*),0]$ (resp. $r(\delta)\in[0,r(\delta^*)]$) for $\delta\in\inter$. The two cases are gathered as: 
	$r(\delta) \in \frac12 r(\delta^*) \pm \frac12 |r(\delta^*)|$ 
	Thus, $\exists \epsilon\in\inter$, $r(\delta)=f(x) - (\bmid{y} + \brad{y} \delta) = \frac12 r(\delta^*) + \frac12 |r(\delta^*)| \epsilon$, 
	as $x=\bmid{x} + \brad{x} \delta$ by definition of $\mu$ as in Lemma~\ref{lem:unitmap}.
	Then, $g(\delta,\epsilon)$ and the proof follow from the last equality. 
\begin{cor} \label{cor:lininclmeth}
	The s-function $[\mu^{-1}(\delta); g(\delta,\epsilon)]$ where $\delta$ and $\epsilon$ refer to symbols of type (unit) interval is a continuous s-zonotope since $\mu^{-1}$ and $g$ are affine. $\forall x\in[x]$, $[x;f(x)] \in \imset{[\mu^{-1}(\delta); g(\delta,\epsilon)]}_{s,\tau,\iota}$, an e-zonotope usually not reduced to an aligned box due to the dependency of both dimensions on common symbol(s) referred as $\delta$. Moreover, if $x$ is a polynotope, so is $\imset{[\mu^{-1}(\delta); g(\delta,\epsilon)]}_{s,\tau}$.  
\end{cor}
\begin{figure}	
	\begin{center}
		\includegraphics[width=7cm]{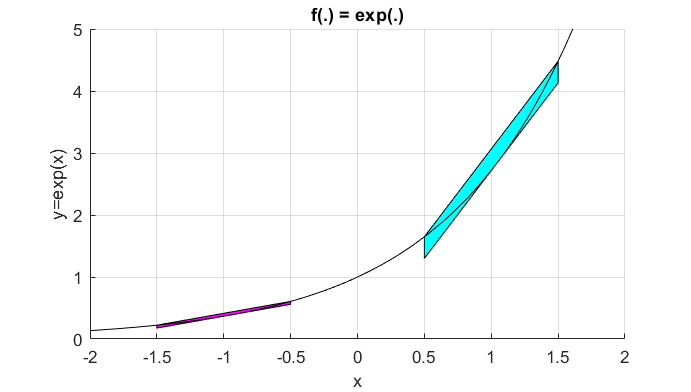}   
	\end{center}
	\caption{Inclusion method of theorem~\ref{thm:lininclmeth} applied to $f(x)=e^x$ on $[x]=-1\pm0.5$ (magenta) and $[x]=+1\pm0.5$ (cyan): plot of e-zonotopes as in corollary~\ref{cor:lininclmeth}. $\bmid{x}+\brad{x}\delta^*=\func{log}(\brad{y}/\brad{x})$.}
	\label{fig:exp}
\end{figure}
An illustrative example with $f(x)=e^x$ is reported in Fig.~\ref{fig:exp} and further remarks are reported hereafter: \\
$a)$ The inclusion proposed in theorem~\ref{thm:lininclmeth} is entirely parameterized by the input domain $[x]$ while not being subject to the arbitrary choice of a point used as reference for linearizing or computing a Taylor expansion. \\
$b)$ $\delta^*$ often has an explicit form, e.g., $\bmid{x}+\brad{x}\delta^*=$ $\func{log}(\brad{y}/\brad{x})$, $\brad{x}/\brad{y}$, $\frac14(\brad{x}/\brad{y})^2$ for $f(x)=e^x$, $\func{log}(x)$, $\sqrt{x}$, respectively. \\ 
$c)$ If $\brad{x}=0$, then the input is a point value and $f(x)=f(\bmid{x})$. This is consistent with the limit $\brad{x}\rightarrow 0$ since the continuity of $f$ gives $\brad{y}\rightarrow 0$, $g_0\rightarrow\bmid{y}$, $g_1\rightarrow0$ and $g_2\rightarrow0$.\\
$d)$ If $f$ is decreasing, then $\bsup{y}<\binf{y}$ and $\brad{y}<0$ in theorem~\ref{thm:lininclmeth}.\\
$e)$ $r(\delta)$ is the remainder term wrt to a (linear) approximation of $f(x)$ which itself (linearly) depends on $x$: $\bmid{y} + \brad{y} \mu(x)$. The purpose of a dependency-preserving inclusion (dpi) is thus achieved, at least for a structural property $\pi$ referring to being linear. Note that polynomial dependencies possibly modeling $x$ (then, $x(.)$ and $\delta(.)=\mu(x(.))$ are polynomials) are readily propagated by a linear enclosing approximation $\tilde{f}(.)$ of  $f(.)$. Indeed, the composition of affine and polynomial functions is still polynomial. Thus, the result in theorem~\ref{thm:lininclmeth} is readily applicable with polynotopes. Moreover, the generic inclusion method in Lemma~\ref{lem:geninclmeth} encompasses polynomial enclosing approximations of non-polynomial functions.

	\subsection{Hybrid: Switching functions} \label{subsec:hybrSWfun}

In the last paragraph ($\S$\ref{subsec:contNLfun}), an inclusion method for functions $f$ satisfying the regularity conditions of being $\mathcal{C}^1$ has been proposed in Theorem~\ref{thm:lininclmeth}. Following the generic inclusion method stated in Lemma~\ref{lem:geninclmeth}, the case of a prototypical $\mathcal{C}^0$ but not $\mathcal{C}^1$ function is considered in this paragraph: the absolute value. The motivation for this is summarized in Table~\ref{table:switchfun} which shows that several useful switching functions can be built by composing basic operators (like $+$, $-$, taking the half) with the absolute value operator $abs(x)=|x|$. Thus, a dependency-preserving inclusion of a prototypical switching function like $abs$ is highly desirable to model and efficiently propagate uncertainties within hybrid dynamical systems, without necessarily requiring costly bisections and/or a specific management of guard conditions.  
\begin{table} 
	\caption{Switching functions expressed from the absolute value operator: $\func{abs}(x)=|x|$ (or from ReLU$^*$).}
	\label{table:switchfun}
	\begin{center}
	\begin{tabular}{lll}
		Function & Notation & Expression with $|.|$ (or $\func{pos}$)\\
		\hline
		Maximum    & $\func{max}(x,y)$ &  $=\frac{x+y}2+\frac{|x-y|}2$ $(=y+\func{pos}(x-y))$ \\
		Minimum    & $\func{min}(x,y)$ &  $=\frac{x+y}2-\frac{|x-y|}2$ $(=x-\func{pos}(x-y))$ \\
		Saturation & $\func{sat}(x,\binf{x},\bsup{x})$ & $=\frac12(\binf{x}+\bsup{x}+|\binf{x}-x|-|x-\bsup{x}|)$ \\ 
		Deadzone   & $\func{dz}(x,\binf{x},\bsup{x})$  & $= x-\func{sat}(x,\binf{x},\bsup{x})$ \\
		ReLU$^*$   & $\func{pos}(x)$   &  $=\func{max}(0,x)=\frac{x+|x|}2$
	\end{tabular}	
	\end{center}
	$^*$Rectifier Linear Unit (remark: $|x|=2\func{pos}(x)-x$).
\end{table}%
\begin{thm}[An inclusion of abs] \label{thm:lininclabs}
	Let $abs: [x] \rightarrow \mathbb{R}, \, x \mapsto y=|x|$ be the restriction of the absolute value on a given interval $[x]=\bmid{x}\pm\brad{x}=[\binf{x}, \bsup{x}] \subset\mathbb{R}$. \\ 
	Case 1: If $\bsup{x}\leq0$, then $abs(.)=-(.)$, \\
	Case 2: If $\binf{x}\geq0$, then $abs(.)=+(.)$, \\
	Case 3: If $|\bmid{x}|<\brad{x}$, then 
	$\widetilde{abs}(.)$ 
	is an inclusion function for $abs(.)$ on $[x]$ with:\\
	\begin{equation}
		\widetilde{abs}(.)=\left(\frac{\bmid{x}}{\brad{x}}\right)(.)+\left(\frac{\brad{x}^2-\bmid{x}^2}{2\brad{x}}\right)(1+\inter).
	\end{equation}
\end{thm}
	\textbf{Proof.}  
	If $0\not\in[x]$ (case 1 or 2), $abs(.)$ is linear and no dedicated inclusion is then required. If $0\in[x]$ (case 3), the inclusion method of theorem~\ref{thm:lininclmeth} is applied step-by-step: Let $\mu$ be the unit range mapping of $[x]$ and $\delta=\mu(x)$. Since $abs(.)$ is only $\mathcal{C}^0$ and convex on $[x]$, but not $\mathcal{C}^1$, the range of the remainder $r(\delta)\leq0$ (still such that $r(-1)=r(+1)=0$) is computed by noticing that its minimum is obtained for $x=\bmid{x}+\brad{x}\delta^*=0$. Then, $\delta^*=-\bmid{x}/\brad{x}$ gives $r(\delta^*)= (\bmid{x}^2-\brad{x}^2)/\brad{x}$ and satisfies $\forall \delta\in\inter$, $r(\delta)\in[r(\delta^*),0]$.  
	It comes $g(\delta,\epsilon)=g_0+g_1\delta+g_2\epsilon$ with
	$g_0 = \frac{\brad{x}^2+\bmid{x}^2}{2\brad{x}}$, 
	$g_1 = \bmid{x}$,
	$g_2 = \frac{\brad{x}^2-\bmid{x}^2}{2\brad{x}}$ ($|\bmid{x}|<\brad{x}$ in case 3).
	Finally, $g(\mu(x),\epsilon)=(\frac{\bmid{x}}{\brad{x}})x+(\frac{\brad{x}^2-\bmid{x}^2}{2\brad{x}})(1+\epsilon)$.
\hspace{\stretch{1}}\\
\rule{0pt}{12pt}%
Corollary~\ref{cor:lininclmeth} still applies to $f=abs$, as a corollary of theorem~\ref{thm:lininclabs} rather than theorem~\ref{thm:lininclmeth}. A dependency-preserving inclusion of $abs(.)$ has been obtained. By extension, dependency-preserving inclusions (dpi) for the switching functions reported in table~\ref{table:switchfun}, among others possibly resulting from functional compositions are also obtained. 
Moreover, the vertical concatenation operator implemented for zonotopes and polynotopes allows to build $n$-dimensional dpi from scalar ones through basic compositions. These can be implemented by using the overloading capability of some object oriented languages, to the benefit of code readability. This feature holds not only for switching functions, but also for  non-linear/non-polynomial ones. This makes polynotopes a relevant tool to compute and analyze mixed uncertainty propagation within non-linear hybrid dynamical systems. Indeed, their polynomial nature, efficiently encoded by combining full and sparse data structures, looks appropriate to model a wide spectrum of non-trivial dependencies, as shown by the functional completeness result given in theorem~\ref{thm:completeness}.

\section{Polynotopic Kalman Filter (PKF)} \label{sec:filtering}

An extension of Kalman Filtering to discrete-time non-linear hybrid dynamical systems is proposed in this section. It is based on polynotopes 
and interpretations related to a set-membership uncertainty paradigm.

Let $x(s)$ be a s-polynotope (\ref{eq:spoly}): $x(s) = \imset{c,R,I,E}_{s,\tau} = c+Rs_I^E$ (syntax). By analogy with zonotopes, its covariation \cite{Combastel2015} is defined as: $\func{cov}(x(s))=RR^T$.
In order to possibly take symbol types and/or the monomial structure into account\footnote{e.g. to weight the relative influence of continuous and discrete symbols/uncertainties on the accuracy criterion further chosen to optimize the mixed-set based state estimates obtained from PKF.}, a covariation weighted by $\varPhi$ (possibly $\varPhi(I,E)$ or any $\varPhi(.)$ depending on known values)   
is introduced as:
\begin{defn}[Weighted covariation]
	Given a symmetric matrix $\varPhi$ ($\varPhi=\varPhi^T$), the weighted covariation of $\spoly{x}=\imset{c,R,\ldots}_{s,\tau(,\iota)}$ (polynotope or zonotope)  is: \\ 
	\rule{25mm}{0mm}$\func{cov}_{\varPhi}(\spoly{x})=R \varPhi R^T$. 
\end{defn}
$x(s)$ formalizes a vector polynomial (s-)function of the symbolic variables in $s_I$. The execution of polynotope operations like sum, linear image, concatenation, reduction, zonotopic hull $\mathcal{Z}{x(s)}$, interval/box hull $\mathcal{B}{x(s)}$, etc mainly work at a syntactic level by manipulating polynomial expressions (e.g. encoded as $(c,R,I,E)$ with sparse $E$) while preserving semantic properties. In particular, inclusion is viewed as a semantic property related to a set-membership interpretation of polynomial functions depending on typed\footnote{Notice that the set-membership interpretation is also related to the types considered under the assumption~\ref{assum:types}.} symbolic variables. From (\ref{eq:epoly}), it comes: $\forall \sigma\in\iota\tau s_{I}, \iota x(\sigma)\in\mathcal{P}x(s)=\imset{c,R,I,E}_{s,\tau,\iota}$ (semantics), where $\iota x(.)$ stands for the interpretation of $x(.)$ as a vector of polynomial mathematical functions ($\mathbb{R}^{dim(I)}\rightarrow\mathbb{R}^{n_x}$) with real coefficients (under assumption~\ref{assum:types}).
Since probability theory is the most commonly used framework for nonlinear filtering, some analogies and exploratory links are briefly outlined as a remark: 
\REVo{ 
\begin{rem} \label{rem:proba}
	A first idea to introduce probability theory in the proposed scheme simply consists in extending the basic symbol types in assumption~\ref{assum:types} to other types like (some class of) random variables defined on a given probability space. This can work in the linear case with symbolic zonotopes and/or independent Gaussian random variables (e.g. see $\S$2.2 and definition~6.1 in \cite{Combastel2019}). Pushing further in such a direction could be an option. Another one may rely on interpreting $\sigma$ (as in definition~\ref{defn:imageset}, (\ref{eq:ezono}) and (\ref{eq:epoly})) as an ``outcome'', the function $\iota x(.)$ as a ``random variable'', $[\iota x]^{-1}(S)$ as an ``event'' related to any set $S$ of output values taken by $\iota x(.)$. A measure $\pi(.)$ of events on the domain $\iota\tau s_{I}$ induced by an interpretation of symbol types would then become some kind of conditional probability wrt $\iota\tau s_I$. The typed symbols $s_I$ would then contribute to define the probability space itself. 
\end{rem}
} 

In the following, no probability measure is considered.
Notations: $\spoly{x} = x(s)$ 
denotes a s-polynotope. $x = \iota x(\sigma) \in \mathbb{R}^{n_x}$ denotes a point evaluation of $\spoly{x}$ obtained for some so-called outcome $\sigma\in\iota\tau s_I$.
Then, $x\in\mathcal{P}\spoly{x}$, the e-polynotope related to $\spoly{x}$.
Also, $x\in\mathcal{Z}\spoly{x}$ (resp. $x\in\mathcal{B}\spoly{x}$) means that $x$ belongs to a zonotopic (resp. interval/box) hull of $\spoly{x}$. 

The state observation (or filtering) problem addressed in this section deals with discrete-time non-linear hybrid dynamical systems modeled as:
\begin{align}
	x_+ & = f(x,u,v), \quad x_0 \in \mathcal{P}\spoly{x}_0, \, v\in\mathcal{P}\spoly{v}, \label{eq:model:state} \\
	0 & = g(x,u,v,y),         \label{eq:model:innov}
\end{align}
where the functions $f(.)$ and $g(.)$ result from the composition of elementary functions and operators for which inclusion preserving polynotope versions are available. In practice, this is not much restrictive since sum, linear image, reduction, concatenation, product are available (see $\S$\ref{subsec:polsfun}) and the modeling tools for nonlinear hybrid systems developed in section~\ref{sec:hybridtools} can be used to that purpose. Then, by overloading\footnote{To the benefit of code readability and maintainability.} these elementary functions and operators with their inclusion preserving polynotopic version, and by applying the same composition, 
inclusion functions with polynotopic inputs and outputs $\tilde{f}(.)$ and $\tilde{g}(.)$ can be obtained for $f(.)$ and $g(.)$, respectively. 
In (\ref{eq:model:state}), the index $+$ refers to the next time step $k+1$ and the current time step $k$ is omitted to simplify the notations, except for the initial state $x_0$ at time $k=0$. $x_0\in\mathbb{R}^{n_x}$ is assumed unknown but bounded by the e-polynotope $\mathcal{P}\spoly{x}_0$ related to a known s-polynotope $\spoly{x}_0$. $x\in\mathbb{R}^{n_x}$, $u\in\mathbb{R}^{n_u}$, $y\in\mathbb{R}^{n_y}$, $v\in\mathbb{R}^{n_v}$ respectively stand for the states, the known (control) inputs, the known measurements, the unknown but bounded uncertainties (state and measurement noises, disturbances, modeling errors, etc) at time $k$. $v$ is assumed bounded by a known polynotope $\mathcal{P}\spoly{v}$. Notice that $u\in\mathcal{P}\spoly{u}=\{u\}$ (singleton) for $\spoly{u}=\imset{u,\emptyset,\emptyset,\emptyset}_{s,\tau}$. Similarly, $y\in\mathcal{P}\spoly{y}$ with $\spoly{y}=\imset{y,\emptyset,\emptyset,\emptyset}_{s,\tau}$. The problem addressed is that of designing a one step-ahead prediction filter (or state observer) minimizing the trace $\func{tr}(.)$ of the (weighted) covariation of a polynotope enclosing the predicted state. 

Filtering is mainly a data fusion process. So, how to merge (vector) sources? Weighting is a usual solution:
\begin{align}
	z_1 \in \mathcal{P}\spoly{z}_1 \wedge z_2 \in \mathcal{P}\spoly{z}_2 \Rightarrow z=G_1z_1+G_2z_2\in\mathcal{P}\spoly{z} \nonumber \\
	\textrm{with}\, \spoly{z}=G_1\spoly{z}_1+G_2\spoly{z}_2. \label{eq:weighting}
\end{align}
Two noticeable ways to particularize (\ref{eq:weighting}) are: \\
$a)$ Taking $z_1=z_2$ under $G_1+G_2=\identity$ gives (\ref{eq:weighting:intersect}) which parameterizes enclosures of a polynotope intersection that could be used to design a state bounding observer: 
\begin{align}
	z \in (\mathcal{P}\spoly{z}_1 \cap \mathcal{P}\spoly{z}_2) & \,\Rightarrow\, z\in\mathcal{P}(G_1\spoly{z}_1+G_2\spoly{z}_2). \label{eq:weighting:intersect} \\
	z \in \mathcal{P}\spoly{z}_1 \,\wedge\, 0 \in \mathcal{P}\spoly{z}_2 & \,\Rightarrow\, z\in\mathcal{P}(\spoly{z}_1-G\spoly{z}_2). \label{eq:weighting:KF}
\end{align}
$b)$ Taking $z_2=0$ and $G_1=\identity$ under $G_2=-G$ gives (\ref{eq:weighting:KF}) which parameterizes an update (or correction) of an initial knowledge $\mathcal{P}\spoly{z}_1$ about $z$ with some other depending knowledge, $\mathcal{P}\spoly{z}_2$, such as the one obtained through some measurements. (\ref{eq:weighting:KF}) thus looks as a prototypical weighting underlying the structure of Kalman Filters. Moreover, in our approach, the symbols possibly shared between $\spoly{z}_1$ and $\spoly{z}_2$ play a key role in the modeling of dependencies. This makes it possible to tune/optimize $G$ so as to maximize uncertainty cancellation\footnote{which is impossible with usual interval arithmetic, subject to the so-called dependency problem.} when computing $\spoly{z}_1-G\spoly{z}_2$. The general idea of Kalman Filters is indeed to optimize the precision of a prediction $\spoly{p}=\spoly{z}_1$ by using a dependent yet complementary source, the innovation $\spoly{e}=\spoly{z}_2$, to update the prediction as $\spoly{p}-G\spoly{e}$ (\ref{eq:pkf:update}).   
Then, the algorithm (\ref{eq:pkf:reduc})$-$(\ref{eq:pkf:update}) implementing an iteration of the proposed Polynotopic Kalman Filter (PKF) follows as in Table~\ref{table:pkf} and theorem~\ref{thm:pkf}. 
\begin{table}
	\caption{PKF iteration: $\spoly{x}_+ = \func{PKF}(\spoly{x}, \spoly{u}, \spoly{v}, \spoly{y}, \tilde{f}, \tilde{g}, \varPhi, q)$:}
	\label{table:pkf}
	\begin{align}
		\bar{\spoly{x}} & = \redop{q}\spoly{x}, \label{eq:pkf:reduc} & reduction \\ 	
		\spoly{p} & = \tilde{f}(\bar{\spoly{x}},\spoly{u},\spoly{v}), & prediction \label{eq:pkf:pred} \\ 
		\spoly{e} & = \tilde{g}(\bar{\spoly{x}},\spoly{u},\spoly{v},\spoly{y}), & innovation \label{eq:pkf:innov} \\
		& \hspace{-8pt}\left< \breve{c}, \left[\begin{array}{c} R_p \\ R_e \end{array}\right], \breve{I}, \breve{E} \right>_{s,\tau} = \left[\begin{array}{c} \spoly{p} \\ \spoly{e} \end{array}\right], & alignment \label{eq:pkf:align} \\
		G & = (R_p \varPhi R_e^T)(R_e \varPhi R_e^T)^{-1}, & optimal\,gain \label{eq:pkf:Gopt} \\
		\spoly{x}_+ & = \spoly{p} - G\spoly{e}. & update \label{eq:pkf:update}
	\end{align}
\end{table}
\begin{thm}[PKF: inclusion and optimal gain] \label{thm:pkf}
	Given a system modeled as in (\ref{eq:model:state})$-$(\ref{eq:model:innov}), the PKF iteration in (\ref{eq:pkf:reduc})$-$(\ref{eq:pkf:update}) (Table~\ref{table:pkf}) satisfies $a)$ and $b)$:\\
	$a)$ $x \in \mathcal{P}\spoly{x} \,\wedge\, v \in \mathcal{P}\spoly{v} \,\Rightarrow\, x_+ \in \mathcal{P}\spoly{x}_+$, \\
	$b)$ Let $G^* = \func{arg}\,\func{min}_G\,\func{tr}(\func{cov}_{\varPhi}(\spoly{x}_+))$. $G^*$ is the optimal gain computed as in (\ref{eq:pkf:Gopt}): $G^*=(R_p \varPhi R_e^T)(R_e \varPhi R_e^T)^{-1}$.
\end{thm}	
	\textbf{Proof.}
	$a)$ : By construction, $\tilde{f}(.)$ and $\tilde{g}(.)$ are inclusion functions for $f(.)$ and $g(.)$. Since the reduction step (\ref{eq:pkf:reduc}) is inclusion preserving, the inclusion property $a)$ is a direct consequence of (\ref{eq:weighting:KF}) with $\spoly{z}_1=\spoly{p}$ and $\spoly{z}_2=\spoly{e}$. \\
	$b) : $
	$\partial_X h(X)$ denoting
	$\partial h(X) / \partial X$,
	if $h(.)$ returns scalar values and $X=[X_{ij}]$ is a matrix, then $\partial_X h(X)=[\partial_{X_{ji}} h(X)]$.
	$X$, $A$, $B$, $C$ being matrices of correct size, 
\REVo{   
	\begin{align}
	\partial_X \func{tr}(AX^TB) & = A^TB^T, \label{eq:dtrAXTB} \\
	\partial_X \func{tr}(AXBX^TC) & = B X^T C A + B^T X^T A^T C^T.  \label{eq:dtrAXBXTC}
	\end{align}
} 
	Let $J(G)=\func{tr}(\func{cov}_{\varPhi}(\spoly{x}_+))$. In (\ref{eq:pkf:align}), $\breve{c}=[c_p; c_e]$ and 
	$[\spoly{p};\spoly{e}]$ is such that\footnote{The polynotope concatenation $[\spoly{p};\spoly{e}]$ gives expressions of $\spoly{p}$ and $\spoly{e}$ such that the generators related to their common monomials (i.e. dependencies) become ``aligned'' in the same columns of the matrices $R_p$ and $R_e$. This is the reason why (\ref{eq:pkf:align}) is called the alignment step.}
	$\spoly{p}=c_p+R_ps_{\breve{I}}^{\breve{E}}$ and $\spoly{e}=c_e+R_es_{\breve{I}}^{\breve{E}}$.
	From (\ref{eq:pkf:update}), $\spoly{x}_+=\imset{c_p-Gc_e, R_p-G R_e,\breve{I},\breve{E}}_{s,\tau}$ and
	$J(G)$
	$=\func{tr}((R_p-G R_e) \varPhi (R_p-G R_e)^T)$
	$=\func{tr}(R_p \varPhi R_p^T)-2\func{tr}(R_p \varPhi R_e^T G^T)+\func{tr}(G R_e \varPhi R_e^T G^T)$.
\REVo{ 
	Using (\ref{eq:dtrAXTB}) and (\ref{eq:dtrAXBXTC}), 
} 
	$\partial_GJ(G) =-2(R_e \varPhi R_p^T)+2(R_e \varPhi R_e^T)G^T$.
	$G^*$ being the value of $G$ such that $\partial_GJ(G)=0$, it comes
	$G^*R_e \varPhi R_e^T = R_p \varPhi R_e^T$ and $G^*=(R_p \varPhi R_e^T)(R_e \varPhi R_e^T)^{-1}$.
\REVo{ 
\begin{thm}[PKF vs. ZKF] \label{thm:pkfvszkf}
	Let consider the particular case of linear functions $f(.)$ and $g(.)$ defined as: \\
	\rule{8mm}{0mm}$f(x,u,v)=Ax+Bu+Ev_p$, \\
	\rule{8mm}{0mm}$g(x,u,v,y)=Cx+Du+Fv_e-y$, \\
	where $v=[v_p;v_e]$ (state noise and measurement noise), and $A, B, C, D, E, F$ are (possibly time-varying) matrices with appropriate dimensions. 
	Only symbols of type (unit) interval are considered and $\varPhi=\identity$.
	Also, let $\tilde{f}(.)=f(.)$, $\tilde{g}(.)=g(.)$,
	$x_0\in\mathcal{P}\spoly{x}_0$ with $\spoly{x}_0=\imset{c_0,R_0,I_0,\identity}_{s,\tau}$ (then, $\mathcal{P}\spoly{x}_0=\mathcal{Z}\spoly{x}_0$ is a zonotope), $v\in\mathcal{P}\spoly{v}$ with $\spoly{v}=\imset{0,\identity,I_v,\identity}_{s,\tau}$ (then, $\mathcal{P}\spoly{v}=\mathcal{Z}\spoly{v}=\mathcal{B}\spoly{v}$ is a unit hypercube). It is also assumed that $I_0$ and all $I_v$'s have no common scalar elements/identifiers which are all unique (then, no symbol being shared between $\spoly{x}_0$ and all the $\spoly{v}$'s, this is in fact an independence assumption). \\	
	Then, 
	PKF computes the same centers $c$ (state point estimates) and generator/shape matrices $R$ as ZKF in \cite{Combastel2015} would do, up to column permutations; all the computed polynotopes are also zonotopes, and the optimal gain $G=AK$ corresponds to the usual Kalman gain $K=\bar{P}C^T(C\bar{P}C^T+FF^T)^{-1}$ with  $\bar{P}=\bar{R}\bar{R}^T$. 	
\end{thm}
	\textbf{Proof.}
	Polynotopes (and zonotopes) being closed under linear transforms, taking  $\tilde{f}(.)=f(.)$ and $\tilde{g}(.)=g(.)$ suffices to preserve inclusion when (\ref{eq:model:state})$-$(\ref{eq:model:innov}) is a discrete-time LTV (or LTI) model. Moreover, all the polynotope exponent matrices equaling $\identity$, (s-)polynotope operations naturally reduce to (s-)zonotope operations, and the generator matrix computed by the considered reduction operator does not depend on the symbolic description. 
	The focus of the proof is first placed on the observer structure and, then, on the optimal gain.
	Observer structure: \\
	(\ref{eq:pkf:reduc}): $\spoly{\bar{x}}=\imset{\bar{c},\bar{R},\bar{I},\identity}_{s,\tau}=\redop{q}\spoly{x}$ where $\bar{c}=c$, \\
	(\ref{eq:pkf:pred}):
	$\spoly{p} = A\bar{\spoly{x}}+B\spoly{u}+E\spoly{v_p}$, \qquad$\,\,$ with $\spoly{v_p}=[\identity, 0]\spoly{v}$, \\
	(\ref{eq:pkf:innov}):
	$\spoly{e} = C\bar{\spoly{x}}+D\spoly{u}+F\spoly{v_e}-\spoly{y}$, \quad with $\spoly{v_e}=[0, \identity]\spoly{v}$, \\
	(\ref{eq:pkf:update}): 
	$\spoly{x}_+ =\spoly{p} - G\spoly{e}$ gives: \\
	$\spoly{x}_+ = (A\bar{\spoly{x}}+B\spoly{u}+E\spoly{v_p})+G(\spoly{y}-(C\bar{\spoly{x}}+D\spoly{u}+F\spoly{v_e}))$, \\
	which corresponds to $(\!(14)\!)$ i.e. the equation (14) in \cite{Combastel2015} where $(v,w)$ stands for $(v_p,v_e)$. Also, just for insight: \\
	$\spoly{x}_+ = (A-GC)\bar{\spoly{x}}+(B-GD)\spoly{u}+[E, -GF]\spoly{v}+G\spoly{y}$. \\
	Keeping in mind that the sum of two generators with the same monomial term (here: with the same symbol) is a classical vector sum, and an horizontal concatenation otherwise (see MLC in $\S$\ref{subsec:linsfun}), the centers $c_*$ and generator matrices $R_*$ of the s-polynotopes (also s-zonotopes since $E_*=\identity$) computed in (\ref{eq:pkf:pred}), (\ref{eq:pkf:innov}) and (\ref{eq:pkf:update}) are\footnote{Up to column permutations with no impact on the interpretation.}: \\
	(\ref{eq:pkf:pred}): $c_p=A\bar{c}+Bu$, \\
	(\ref{eq:pkf:pred}): $R_p=[A\bar{R}, E]$ since $\bar{I} \cap I_{v_p}=\emptyset$, \\
	(\ref{eq:pkf:innov}): $c_e=C\bar{c}+Du-y$, \\
	(\ref{eq:pkf:innov}): $R_e=[C\bar{R}, F]$ since $\bar{I} \cap I_{v_e}=\emptyset$, \\
	(\ref{eq:pkf:update}): $c_+ = c_p - G c_e = (A-GC)\bar{c}+(B-GD)u+Gy$, \\
	(\ref{eq:pkf:update}): $R_+ = R_p - G R_e = [(A-GC)\bar{R}, E, -GF]$, \\
	since $I_{v_p} \cap I_{v_e}=\emptyset$, but note that PKF can take dependent state and measurement noises into account with $\spoly{v}$.\\
	Finally, it can be checked that $c_+$ and $R_+$ exactly coincide with $(\!(15)\!)$ and $(\!(16)\!)$, respectively, so proving that PKF reduces to the same observer structure as ZKF under the specific assumptions of theorem~\ref{thm:pkfvszkf}. \\ 
	Optimal gain: Let $\bar{P}=\bar{R}\bar{R}^T$ ($=\func{cov}_{\varPhi}(\spoly{\bar{x}})$, $\varPhi=\identity$). Respectively substituting $[A\bar{R}, E, 0]$ and $[C\bar{R}, 0, F]$ for $R_p$ and $R_e$ in (\ref{eq:pkf:Gopt}) 
	gives $G=AK$ with $K=\bar{P}C^T(C\bar{P}C^T+FF^T)^{-1}$. 
	Then, it can be checked that the optimal observer gain is the same as in $(\!(21)\!)- (\!(22)\!)$. 
} 
\begin{rem} [PKF vs. KF] \label{thm:pkfvskf}
	Theorem~\ref{thm:pkfvszkf} (PKF vs. ZKF) can be combined with Theorem~7 (ZKF vs. KF) in \cite{Combastel2015} to make a further bridge between  set-membership and stochastic paradigms. In particular, this gives the conditions under which PKF covariations and KF covariances coincide, as well as the state point estimates.  
\end{rem}
Based on modeling tools for nonlinear hybrid systems developed in the proposed approach, a compositional implementation of advanced reachability and filtering algorithms preserving inclusion is made possible by using operator overloading. This is exemplified with the Polynotopic Kalman Filter (PKF) proposed in this section.

\section{Numerical Examples} \label{sec:numex}

	\subsection{Discrete: Adder} \label{subsec:discAdder}

The first example illustrates some connection with basic digital circuit design. The s-polynotopes (i.e. polynomial s-functions) resulting from the multi-affine decomposition of $n$ bits binary adders only made of \emph{nand} gates are compared depending on the type of symbol(ic variable)s used: signed or boolean as explained in $\S$\ref{subsec:discSBfun}.%

\begin{table} 
	\caption{Algorithm of functions building Half (H), Full 1 bit (F), and Full $n$ bits (N) adder with \emph{nand} gates ($\neg x \leftrightarrow x \nnd x$).}
	\label{table:adderarchi}
	\begin{tabular}{l}
		\hline
		$H: (a,b) \mapsto (s,c) :$ \hspace{\stretch{1}} Half-adder with 5 nand gates \\ 
		$t_1 \eqa a   \nnd b  $, 
		$t_2 \eqa a   \nnd t_1$, 
		$t_3 \eqa t_1 \nnd b  $, 
		$s   \eqa t_2 \nnd t_3$,
		$c   \eqa t_1 \nnd t_1$. \\
		\hline 
		$F: (a,b,c_{in}) \mapsto (s,c_{out}) :$ \hspace{\stretch{1}} Full 1 bit adder with carry \\
		$(r,c_1) \eqa H(a,b)$,
		$(s,c_2) \eqa H(r,c_{in})$,
		$c_{out} \eqa \neg c_1 \nnd \neg c_2$. \\
		\hline 
		$N: (A,B,c) \mapsto (S,c) :$ \hspace{\stretch{1}} Full $n$ bits adder with carry\\
		$\textrm{for}\,i \eqa 1\ldots n(A), \, (S(i),c) \eqa F(A(i),B(i),c)$. \\
		\hline
	\end{tabular}
\end{table}%
The binary adders architecture is described in Table~\ref{table:adderarchi} where $S(i)$ refers to the projection of the s-polynotope $S$ along the $i$th dimension, $i=1,\ldots,n(S)$. For an $n$ bits adder, $A=s_{1:n}$ and $B=s_{(n+1):2n}$ each refer to a vector of $n$ (either signed or boolean) symbolic variables representing the (unknown) bits encoding two integer operands. The e-zonotope (or e-polynotope) related to $A$ is the set of the $2^n$ possible input values i.e. $\{-1,+1\}^n$ (resp. $\{0,1\}^n$) in the signed (resp. boolean) case. Idem for $B$. These sets are never computed explicitly: they only describe the set-valued interpretation of semi-symbolic calculi based on the $(c,R,I,E)$ data structure used to represent s-polynotope objects. Then, building the architecture of an $n$ bits adder by computing $(S,c)=N(A,B)$ as in table~\ref{table:adderarchi} with s-polynotope overloaded operators results in the s-polynotopes $S$ (sum result) and $c$ (output carry) of dimension $n$ and $1$, respectively. 
$S(i)$ is a polynomial with scalar coefficients giving the expression of the $i$th bit of $S$ as a function of the input bits/symbols in $A$, $B$ and an input carry.
A full $n$-bits adder has $2n+1$ (binary) inputs and $n+1$ (binary) outputs. The s-polynotope $\tilde{S}=[S;c]$ gathering the sum result and the output carry is thus given by $\tilde{S}=\imset{\tilde{c},\tilde{R},\tilde{I},\tilde{E}}_{s,\tau}$ with $\tilde{c}\in\mathbb{R}^{n+1}$, $\tilde{R}\in\mathbb{R}^{(n+1)\times m(n)}$, $\tilde{I}\in\mathbb{N}^{2n+1}$, $\tilde{E}\in\mathbb{N}^{(2n+1)\times m(n)}$.
The number of (distinct) generators/monomials in $\tilde{S}$, including the center/constant term, is $1+m(n)$. 
\begin{table}
	\caption{Number of distinct generators/monomials of s-polynotopes representing the multi-affine decomposition of an $n$-bits adder (as in table~\ref{table:adderarchi}) with either signed or boolean symbol(ic variable)s. The computation time is given in seconds (s).}
	\label{table:adderresults}
	\begin{center}
		\begin{tabular}{l@{\rule{2mm}{0mm}}c@{\rule{2mm}{0mm}}c@{\rule{2mm}{0mm}}c@{\rule{2mm}{0mm}}c@{\rule{2mm}{0mm}}c@{\rule{2mm}{0mm}}c@{\rule{2mm}{0mm}}c@{\rule{2mm}{0mm}}c}
			$n$        & 1    & 2    & 3    & 4    & 5    & 6    & 7   & 8   \\
			\hline
			Signed     & 5    & 11   & 23   & 47   & 95   & 191  & 383 & 767 \\
			\qquad (s) & 0.01 & 0.02 & 0.03 & 0.06 & 0.15 & 0.36 & 1.9 & 8.3 \\
			\hline
			Boolean    & 8    & 23   & 65   & 188  & 554  & 1649 &  -  &  -  \\
			\qquad (s) & 0.01 & 0.01 & 0.03 & 0.2  & 2.3  & 29   &  -  &  -  \\
			\hline
		\end{tabular}	
	\end{center}
\end{table}
This number is reported in Table~\ref{table:adderresults} depending on the number $n$ of bits of the adder and the symbol types: either signed or boolean.
An unexpected yet interesting 
result is obtained: The number of distinct monomials required to describe the full architecture of an $n$-bits adder is much smaller and more scalable using signed rather than boolean symbols. 

This example of a full $n$-bits adder shows the ability of s-polynotopes to describe and manipulate purely discrete expressions yielding non trivial relations and dependencies between inputs and outputs. 
Though requiring further studies, the automatic reduction if such relations could help to struggle against combinatorial explosion by gathering into continuous domains the influence of many bits/signs having a small influence on a given criterion, while keeping trace of how the most significant ones influence that criterion.
Moreover, the polynomial representation benefits from useful simplifications made possible by dealing with typed symbols.

\subsection{Continuous: Van-Der-Pol oscillator} \label{subsec:contVDPoscil}

In order to illustrate reachability on continuous domains and compare the results with \cite{Kochdumper2019}, a Van-Der-Pol oscillator taken from \cite{Immler2018} is considered:
\begin{align}
	\dot{x}_1 & = x_2, \nonumber \\
	\dot{x}_2 & = (1-x_1^2)x_2-x_1. \nonumber
\end{align}
\begin{figure}
	\begin{center}
		\includegraphics[width=6cm]{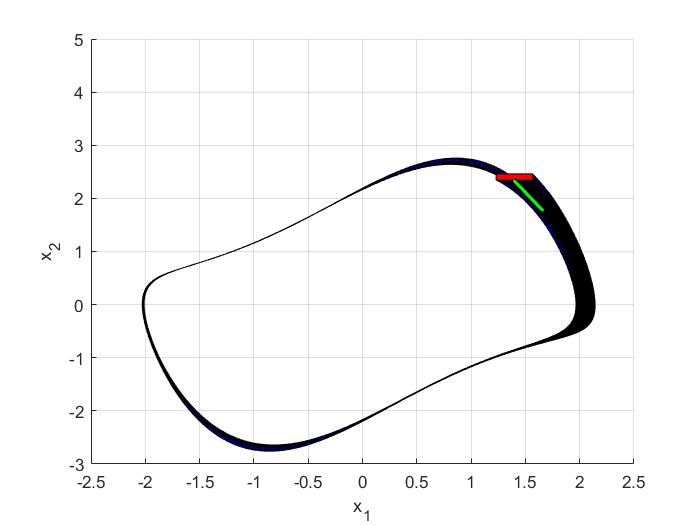}
	\end{center}
	\caption{Reachability result obtained on the Van-Der-Pol oscillator with continuous polynotopes (the plot results from zonotopic enclosures $\mathcal{Z}\spoly{x}$ at each time step).}
	\label{fig:vdp}
\end{figure}%
The initial state set is $\mathcal{P}\spoly{x}_0 =  \mathcal{P}[\spoly{x}_{1,0}; \spoly{x}_{2,0}] = [[1.23, 1.57];$ $[2.34, 2.46]]$ as shown by the red box in Fig.~\ref{fig:vdp}. $1360$ iterations based on an Euler sampling with step size $h=0.005$ are computed in $\SI{9.8}{\second}$ with continuous polynotopes under Matlab running on a 1.8~GHz Core i5 processor with 8~Go RAM. The zonotopic enclosure $\mathcal{Z}\spoly{x}_{1360}$ of the final polynotope (at $t=\SI{6.8}{\second}$) is the green set in Fig.~\ref{fig:vdp}.
At each iteration, the (polynotopic version of the) reduction operator $\redop{q}$ from \cite{Combastel2019} with $q=50$ is used to: $a)$ reduce the square $\spoly{x}_1^2$, $b)$ reduce the product $(\spoly{x}_1^2)\spoly{x}_2$, $c)$ reduce $\spoly{x}$.
As expected, the reachability result shown in Fig.~\ref{fig:vdp} is close to the one obtained with sparse polynomial zonotopes (spz) in the Figure~6 of \cite{Kochdumper2019}, where a comparison with other methods is conducted. Thus, continuous polynotopes also outperform zonotopes and quadratic zonotopes on this example, which illustrates the interest in dealing with polynomial dependencies to propagate continuous domains within nonlinear dynamics.

	\subsection{Hybrid: Traffic network} \label{subsec:HybrTrafnet}

In order to illustrate reachability for dynamics defined with switching functions like $\func{min}$ (see $\S$\ref{subsec:hybrSWfun} and table~\ref{table:switchfun}) and compare the results with those reported for the TIRA toolbox in \cite{Meyer2019}, the model of a $3$-link traffic network representing a \emph{diverge} junction is considered:
\begin{align}
	\dot{x}_1 & = -k(x)/T+p, \nonumber \\
	\dot{x}_2 & = k(x)/2 - \func{min}(c,vx_2), \nonumber \\
	\dot{x}_3 & = k(x)/2 - \func{min}(c,vx_3), \nonumber \\
		\textrm{where} \, & k(x) = \func{min}(c, vx_1, 2w(\bar{x}-x_2), 2w(\bar{x}-x_3)). \nonumber
\end{align}
$\func{min}(.,.,.,.)$ is implemented as $\func{min}(\func{min}(.,.),$ $\func{min}(.,.))$. The state $x\in\mathbb{R}^3$ is the vehicle density on each link. $p\in[4/3,2]$ is the constant but uncertain vehicle inflow to link 1. Notice that the constant nature of this uncertainty is naturally handled by the proposed symbolic approach (no new symbol at each time for $\spoly{p}$). As in \cite{Meyer2019}, the known parameters of the network are $[T, c, v, \bar{x}, w]=[30, 40, 0.5, 320, 1/6]$. The initial state set is $\mathcal{P}\spoly{x}_0 = [[150, 200]; [180, 300]; [100,220]]$. An Euler sampling with step size $h=1$ and final time $t_f=30$ is considered. The reduction $\spoly{x}=\redop{q}\spoly{x}$ is applied at each iteration with $q=20$. 
\begin{figure}
	\begin{center}
		\includegraphics[width=6cm]{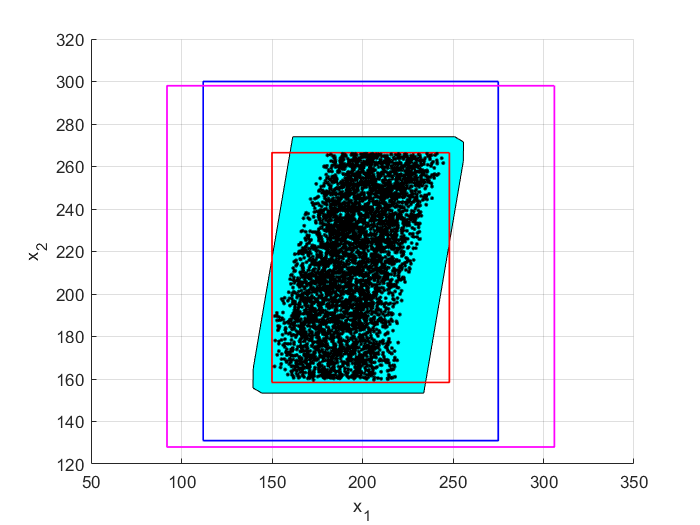}
	\end{center}
	\caption{Reachability result obtained with polynotopes (cyan) for the model of a $3$-link traffic network representing a \emph{diverge} junction. Comparison with the results reported in Figure~2 in \protect\cite{Meyer2019}: methods C/GB (blue), SDMM-IA (magenta), MM and SDMM-S/F (red); Monte-Carlo simulations (black dots).}
	\label{fig:tira}
\end{figure}%
Then, $\SI{0.26}{\second}$ were required to compute the polynotope $\mathcal{P}\spoly{x}_{t_f}$ reported in cyan in Fig.~\ref{fig:tira}.
For the sake of a first comparison, 
the results in the Figure~2 in \cite{Meyer2019} are also reported in Fig.~\ref{fig:tira} : C/GB (Contraction/Growth Bound), MM (Mixed Monotonicity), SDMM-IA (Sampled Data MM-Interval arithmetic), SDMM-S/F(Sampled Data MM-Sampling/Falsification). $\mathcal{P}\spoly{x}_{t_f}$ looks competitive wrt to the results obtained with TIRA (red box in Fig.~\ref{fig:tira}). Moreover, the computed polynotope captures the orientation of the ``black cloud of sample successors'' obtained from $5000$ Monte-Carlo simulations and also reported in Fig.~\ref{fig:tira}. This illustrates the ability of the proposed scheme to maintain dependency links while propagating uncertainties through hybrid dynamics modeled with switching functions.

	\subsection{Reachability and Filtering: Lotka-Volterra} \label{subsec:RFlotka}

A non-linear non-autonomous prey-predator model resulting from the discretization of a modified continuous-time Lotka-Volterra model illustrates $a)$ the computation of reachable sets based on a mixed-encoding ($\S$\ref{subsec:mxenc}) of the initial state set, and $b)$ Polynotopic Kalman Filtering (PKF) as developed in section $\S$\ref{sec:filtering}. 
The modified continuous-time Lotka-Volterra model is $\dot{x}=\mathsf{f}(x,u)$ with $x\in\mathbb{R}^2$, $u\in\mathbb{R}$, $(a, b, c, d) = (2, 0.4, 1, 0.1)$, and
$\mathsf{f}(x,u) = [a x_1 - b x_1 x_2; \, -c x_2 + d x_1 x_2 + u]$.

		\subsubsection{Reachability with a mixed-encoding of state sets} \label{subsubsec:Rlotka}

\begin{figure}
	\rule{-4mm}{0mm}\includegraphics[width=9cm]{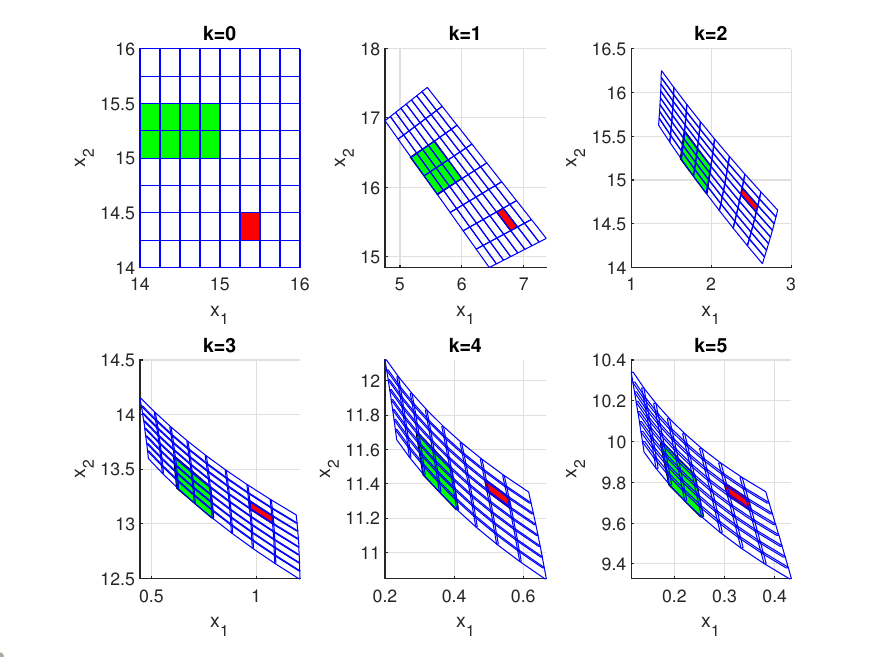}
	\caption{Reachable sets resulting from a 3-level signed-interval mixed encoding of the initial states of a Lotka-Volterra model. The mixed polynotope computed at each time $k$ characterizes (overlapping) outer approximations of the propagation of each of the 64 cells (red) paving the initial state set \emph{with no bissection}. This also works with cell groups (green).}
	\label{fig:lotkas}
\end{figure}

A mixed-encoding of the initial state set is first considered and further propagated using mixed polynotope computations within the non-linear dynamics of the discretized Lotka-Volterra model. \\ 
More precisely, following the definition~\ref{defn:mxenc} and the corollary~\ref{cor:mxenc_int}, a 3-level signed-interval mixed encoding of the interval $15\pm1$ is taken as polynotopic initial state set for both $x_{1,0}$ and $x_{2,0}$ (i.e. $x_1$ and $x_2$ at $t=kh=0$): 
\begin{align}
	& \spoly{x}_{1,0}=15+1.Z_s^3([!(3,\mathtt{s});!(1,\mathtt{i})]), \label{eq:lotkas:x10} \\
	& \spoly{x}_{2,0}=15+1.Z_s^3([!(3,\mathtt{s});!(1,\mathtt{i})]), \label{eq:lotkas:x20} \\
	& \mathcal{P}\spoly{x}_0 =  \mathcal{P}[\spoly{x}_{1,0}; \spoly{x}_{2,0}] = [[14, 16]; [14, 16]]. \label{eq:lotkas:Px0} 
\end{align}
Each occurrence of $!(3,\mathtt{s})$ (resp. $!(1,\mathtt{i})$) calls USP (see $\S$\ref{subsec:TypedSymbols}) which returns 3 (resp. 1) unique identifiers of symbols of type signed (resp. unit interval). Thus, $\spoly{x}_{1,0}$ and $\spoly{x}_{2,0}$ are independent since they share no common symbol. Note that the symbol types are compatible with the definition~\ref{defn:mxenc} of $Z_s^g$. $g$ refers to the granularity level of the mixed-encoding. $g=3$ means that $3$ signed symbols are used to hierarchically decompose the range $15\pm1$ into $2^3=8$ sub-intervals. The coverage of the continuous domain $15\pm1$ is then ensured by the remainder term modeled by the symbol of type unit interval uniquely identified by $!(1,\mathtt{i})$.
For instance, let $I=[!(3,\mathtt{s});!(1,\mathtt{i})]$ and $\spoly{x}_{1,0}=15+1.Z_s^3(I)$ as in (\ref{eq:lotkas:x10}). Then, $\spoly{x}_{1,0} = \imset{15, [\frac12, \frac14, \frac18, \frac18], I, 1}_{s,\tau} = \frac12 s_{I_1} + \frac14 s_{I_2} + \frac18 s_{I_3} + \frac18 s_{I_4}$, where the 3 symbols $s_{I_{1:3}}$ are of type signed i.e. $\iota s_{I_{1:3}} \in \{-1,+1\}^3$ and the symbol $s_{I_4}$ is of type (unit) interval i.e. $\iota s_{I_4} \in [-1,+1]$, so that $\spoly{x}_{1,0}$ also writes as:
\begin{align}
	& \spoly{x}_{1,0} = \frac12 \sign_{I_1} + \frac14 \sign_{I_2} + \frac18 \sign_{I_3} + \frac18 \inter_{I_4}. \label{eq:lotkas:x10bis} \\
	& \spoly{x}_{2,0} = \frac12 \sign_{J_1} + \frac14 \sign_{J_2} + \frac18 \sign_{J_3} + \frac18 \inter_{J_4}. \label{eq:lotkas:x20bis} \
\end{align}
$\spoly{x}_{2,0}$ is obtained analogously from $J=[!(3,\mathtt{s});!(1,\mathtt{i})]$ ($I \cap J = \emptyset$) and $\spoly{x}_{2,0}=15+1.Z_s^3(I)$ as in (\ref{eq:lotkas:x20}).
The e-polynotope (or e-zonotope) related to the s-polynotope $\spoly{x}_{1,0}$ (or s-zonotope since $E=\identity$) is thus $\mathcal{P}\spoly{x}_{1,0}=\mathcal{Z}\spoly{x}_{1,0}=15\pm1$ (corollary~\ref{cor:mxenc_ezono}). The independence of $\spoly{x}_{1,0}$ and $\spoly{x}_{2,0}$ coming from $I \cap J = \emptyset$ gives (\ref{eq:lotkas:Px0}). 
The polynotope $\spoly{x}_0 = [\spoly{x}_{1,0}; \spoly{x}_{2,0}]$ contains all the information required to decompose the initial state set $[[14, 16]; [14, 16]]$ into a paving as in Fig.~\ref{fig:lotkas} for $k=0$.
Moreover, assigning values $+1$ or $-1$ to evaluate some of (or all) the signed symbols in (\ref{eq:lotkas:x10bis})$-$(\ref{eq:lotkas:x20bis}) makes it possible to query about the range covered under some conditional values of signed symbols. This feature makes it possible to trace how each cell and/or cell groups within the ``implicit paving'' of the initial state set will propagate. It is worth underlining that this can be achieved \emph{without any bissection}, only through the polynomial computations implementing the basic polynotope operators. 

For the sake of illustration, an Euler sampling of the Lotka-Volterra model is considered: $x_+ = x + \mathsf{f}(x,0)h$, where the time index $k$ is omitted, $x_+$ stands for $x_{k+1}$, and the step size is $h=\SI{0.15}{\second}$. The reduction operator $\redop{50}$ is applied at each iteration. The reachability analysis reported in Fig.~\ref{fig:lotkas} results from $N=5$ iterations starting from the initial state (\ref{eq:lotkas:x10})$-$(\ref{eq:lotkas:x20}) satisfying (\ref{eq:lotkas:Px0}). 
Let $[+-+]$ be a short notation for $[+1;-1;+1]$ (also applying for other sign combinations). Let $\spoly{x}|(s_I=v)$ denote the s-polynotope obtained by substituting in $\spoly{x}$ the expressions in $v$ for the symbolic variables in $s_I$. Unless $v$ depends on some symbols in $s_I$, $\spoly{x}|(s_I=v)$ does not depend anymore on $s_I$. The related e-polynotope (resp. e-zonotopic enclosure) is $\mathcal{P}\spoly{x}|(s_I=v)$ (resp. $\mathcal{Z}\spoly{x}|(s_I=v)$). Considering the vectors of unique symbol identifiers $I$ and $J$ as in (\ref{eq:lotkas:x10bis})$-$(\ref{eq:lotkas:x20bis}), the red (resp. green) cell at $k=0$ in Fig.~\ref{fig:lotkas} corresponds to $\mathcal{P}\spoly{x}_0|(s_{I_{1:3}}=[+-+], s_{J_{1:3}}=[--+])$ (resp. $\mathcal{P}\spoly{x}_0|(s_{I_{1}}=[-], s_{J_{1:2}}=[+-])$). Then, the single s-polynotope $\spoly{x}_k$ computed at each iteration $k$ contains all the information required to obtain the related subplot in Fig.~\ref{fig:lotkas}. In particular, the red (resp. green) zonotopic sets are $\mathcal{Z}\spoly{x}_k|(s_{I_{1:3}}=[+-+], s_{J_{1:3}}=[--+])$ (resp. $\mathcal{Z}\spoly{x}_k|(s_{I_{1}}=[-], s_{J_{1:2}}=[+-])$) for $k=1,\ldots,5$. These zonotopic enclosures are guaranteed to enclose the set of states reached from the initial red cell (resp. green cells) by iterating the sampled non-linear dynamics. Moreover, the initial ``implicit paving'' gradually leads to possibly overlapping cells (see the blue borders in Fig.~\ref{fig:lotkas}) since the complexity of the polynotope computed at each iteration is reduced to a finite number ($50$) of generators.
\begin{figure}
	\begin{center}
		\includegraphics[width=6.5cm]{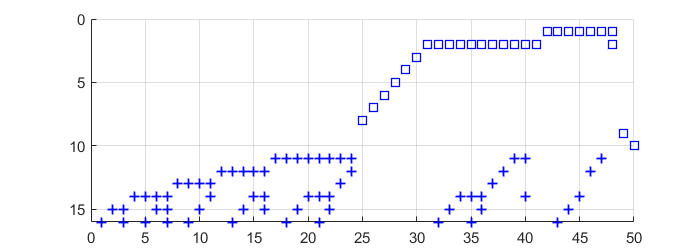}   
	\end{center}
	\caption{Sparse structure of the exponent matrix $E$ of $\spoly{x}$ at $k=5$: 16 symbolic variables (interval: $\square$ marks, signed: $+$ marks), 50 monomials/generators, $90$ non-zero elements.}
	\label{fig:lotkasEsp}
\end{figure}\\
The sparse structure of the exponent matrix $E$ of $\spoly{x}$ at $k=5$ is given in Fig.~\ref{fig:lotkasEsp}. The monomials involve $16$ symbolic variables ($10$ of type unit interval: square marks, $6$ of type signed: $+$ marks). There are $90$ non-zeros elements. The maximum degree is $3$. As expected from the propagation of the initial mixed-encoding of states through a non-linear dynamic, several mixed monomials can be observed, that is, monomials involving both continuous and discrete symbolic variables.

This example of reachability with a mixed-encoding of state sets illustrates the ability of mixed polynotopes to trace the propagation of a significant number of hierarchically organized cells within non-trivial dynamics. Bissections have been avoided by dealing with polynomial dependencies between symbolic variables of different types combining continuous and discrete value domains.    
Moreover, the ability to explicitly characterize the overlapping between cells forming a partition of the initial state set paves the way for efficient 
symbolic abstraction techniques.
		
		\subsubsection{Non-linear and mixed filtering with PKF} \label{subsubsec:Flotka}

The Polynotopic Kalman Fiter (PKF) developed in section~$\S$\ref{sec:filtering} is applied to the modified Lotka-Volterra dynamic without and with a mixed encoding of states:
\begin{align}
	x_+ & = x + \mathsf{f}(x,u)h + E\bar{v}, \label{eq:lotkaf:pred} \\
	y   & = x_1 + F\bar{w}. \label{eq:lotkaf:meas}
\end{align}
The prediction model (\ref{eq:lotkaf:pred}) (resp. measurement equation (\ref{eq:lotkaf:meas})) correspond to (\ref{eq:model:state}) (resp. (\ref{eq:model:innov})) in the formulation of PKF with $v=[\bar{v};\bar{w}]$ and $\varPhi=\identity$. The step size is $h=0.04$ and $k\in\{0,\ldots,N\}\subset\mathbb{N}$ with $N=750$ iterations. The initial and final times are $t_0=0$ and $t_f=Nh=30$. The input is $u=2$ for $t\in[10,20[$ i.e. $250\leq k < 500$, and $u=0$ otherwise. $E=3.10^{-3}\identity$, $F=1.5$. The state and measurement noises are assumed to be bounded as $\bar{v}\in[-1,+1]^2$ and $\bar{w}\in[-1,+1]$ i.e. $\mathcal{P}\spoly{v}=[-1,+1]^3$. The initial state set is assumed to be bounded by $\mathcal{P}\spoly{x}_0=[[5,25]; [5,25]]\subset\mathbb{R}^2$. These bounds are obtained from:
\begin{align}
	& \spoly{x}_{1,0}=15+10.Z_s^g([!(g,\mathtt{s});!(1,\mathtt{i})]), \nonumber \\ 
	& \spoly{x}_{2,0}=15+10.Z_s^g([!(g,\mathtt{s});!(1,\mathtt{i})]), \nonumber \\ 
	& \spoly{v}=[Z_s^0(!(1,\mathtt{i})); Z_s^0(!(1,\mathtt{i})); Z_s^g([!(g,\mathtt{s});!(1,\mathtt{i})])], \nonumber 
\end{align}
where $g$ stands for the granularity level of a mixed encoding of the initial states and the measurement noise at each sample time $k$. Two cases are considered: \\
$i)$ $g=0$ corresponds to a purely continuous case (solid lines in Fig.~\ref{fig:lotkaf}) with only symbols of type unit interval. \\
$ii)$ $g=2$ corresponds to a mixed case (dashed lines in Fig.~\ref{fig:lotkaf}) involving symbols of different types (signed and interval) in mixed polynotope computations. \\
At each iteration of PKF, the reduction operator $\redop{q}$ with $q=50$ or $q=100$ is used to: $a)$ implement the reduction of $\spoly{x}$ as in (\ref{eq:pkf:reduc}), $b)$ reduce the product $\spoly{x}_1\spoly{x}_2$ in $\mathsf{f}(\spoly{x},\spoly{u})$; this is the only (inclusion preserving) difference between $f$ (\ref{eq:model:state}) and $\tilde{f}$ (\ref{eq:pkf:pred}) in this example. Notice also that $g$ (\ref{eq:model:innov}) equals $\tilde{g}$ (\ref{eq:pkf:innov}) since the measurement/innovation equation (\ref{eq:lotkaf:meas}) is linear. 
The simulation of the ``true'' system is obtained from $x_0=[22;8]$ using Heun's method. 
Consistently with (\ref{eq:lotkaf:meas}), only the first state $x_1\in\mathbb{R}$ is measured at each sample time $k$, and the main purpose of PKF is to estimate state bounds $\mathcal{B}\spoly{x}$ for the state $x\in\mathbb{R}^2$ while minimizing the (predicted) polynotope covariation trace.
\begin{table}
\REVo{ 
	\caption{Computation times for $750$ iterations of PKF in seconds (s). Cases: continuous ($g=0$) vs. mixed ($g=2$), and $\redop{50}$ vs. $\redop{100}$.}
	\label{table:lotkaf}
	\begin{center}
		\begin{tabular}{lcc}
				    & $g=0$   &  $g=2$  \\
			\hline
			$q=50$  &  $\SI{3.4}{\second}$ &  $\SI{3.8}{\second}$ \\
			$q=100$ & $\SI{13.9}{\second}$ & $\SI{14.8}{\second}$ \\
			\hline
		\end{tabular}	
	\end{center}
} 
\end{table}
\begin{figure}
\REVo{ 
	\begin{center}
			\includegraphics[width=4cm]{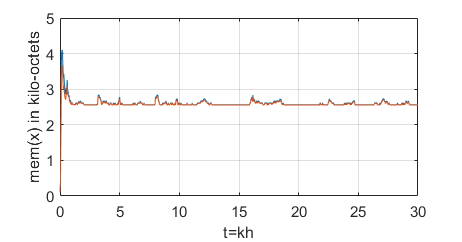} \includegraphics[width=4cm]{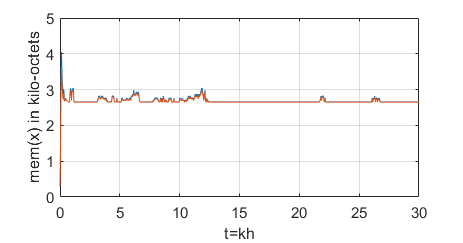} \\
			\includegraphics[width=4cm]{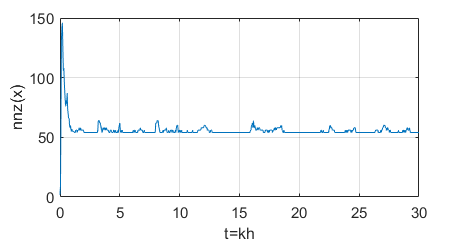} \includegraphics[width=4cm]{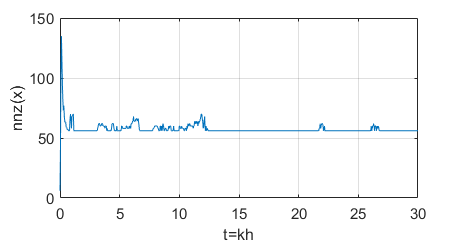}	
	\end{center}
	\caption{Memory footprint of the polynotopic state estimate $\spoly x$ with PKF: Evolution in time under $\redop{50}$. Left: Continuous ($g=0$). Right: Mixed ($g=2$). Top: Memory size $mem(\spoly x)$ in kilo-octets occupied by the object $\spoly x$, depending on whether the sparse exponent matrix $E$ of $\spoly x$ is transposed (red) or not (blue). Bottom: Number of non-zero elements in $E$.}
	\label{fig:lotkafmem}
} 
\end{figure}
\begin{figure}
	\begin{center}
		\includegraphics[width=7cm]{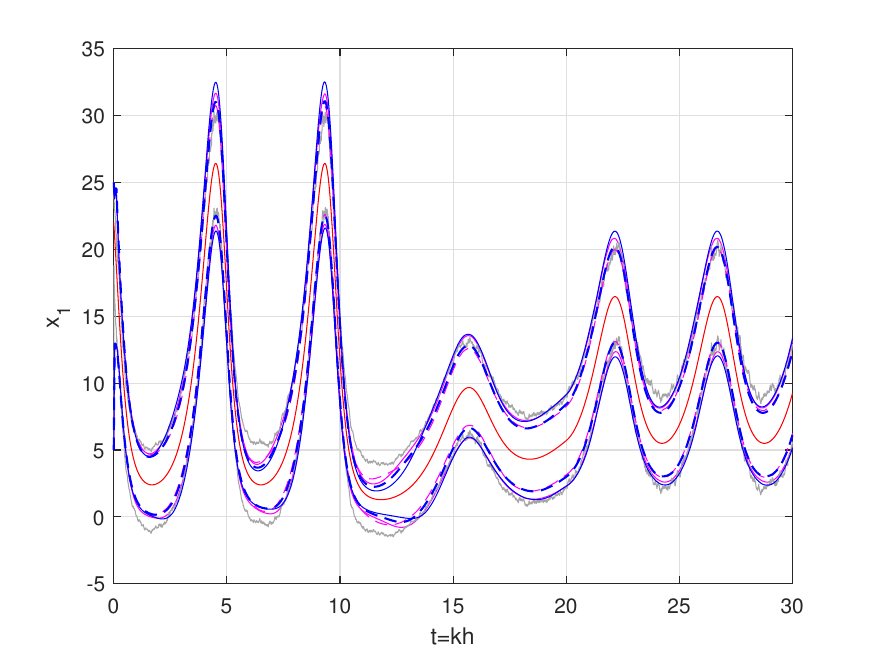} \\
		\includegraphics[width=7cm]{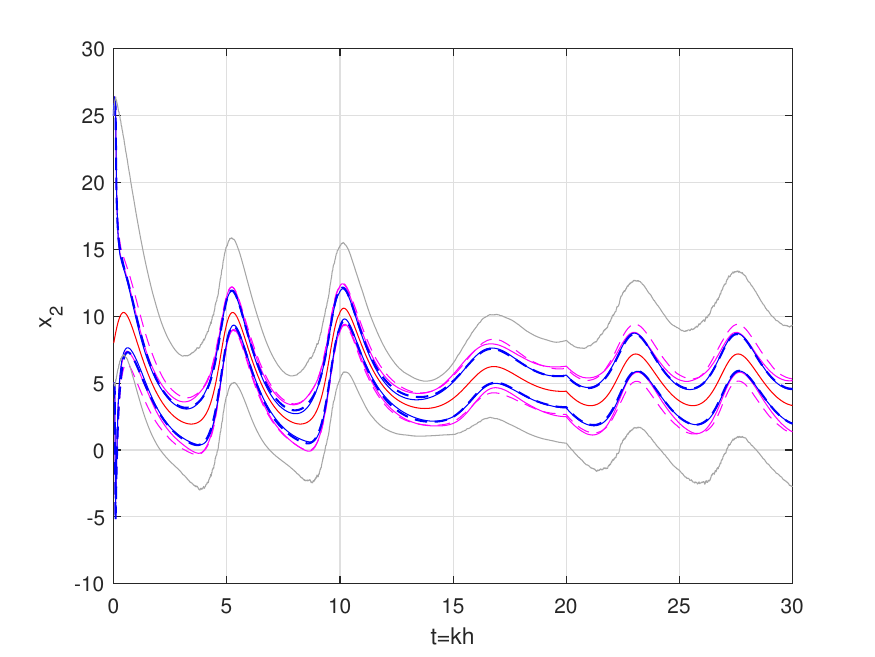}
	\end{center}
	\caption{Estimated state bounds vs. ``true'' values (red) of $x_1$ (top) and $x_2$ (bottom) with the Lotka-Volterra model: PKF with continuous (solid lines) or mixed (dashed lines) polynotopes, and $\redop{50}$ (magenta) or $\redop{100}$ as reduction operator. Comparison with EZGKF in a bounded-error setting (grey).}
	\label{fig:lotkaf}
\end{figure}%
\REVo{%
The table~\ref{table:lotkaf} reports the computation times with a Matlab implementation%
}%
: The mixed encoding does not increase very significantly the computation time in spite of the number of discrete configurations, contrary to the number of generators.
The evolution in time of the memory footprint of $\spoly{x}$ with $\redop{50}$ is also reported in Fig.~\ref{fig:lotkafmem}: once again, only a slight increase is observed in the mixed case ($g=2$) compared to the continuous one ($g=0$). 
The simulation results reported in Fig.~\ref{fig:lotkaf} show a significant improvement of accuracy compared to EZGKF in a purely bounded-error setting which requires around $\SI{2.5}{\second}$ with $q=200$ generators as in \cite{Combastel2016}. In particular, PKF shows an enhanced ability to reconstruct $x_2$ from noisy measurements of $x_1$. Mixed encoding tends to give results with increased accuracy, especially for $x_1$, provided the number of generators is sufficient. Meanwhile, the maximum degree of computed polytopes is decreased from $6$ (resp. $7$) in the continuous case $g=0$ with $q=50$ (resp. $q=100$) to only $4$ in both mixed cases i.e. $g=2$ with $q=50$ or $100$. This is consistent with the reduced size of remainder intervals for $g=2$ and the influence of rewriting rules, in particular the inclusion neutral rule (\ref{eq:rrspow}). This illustrates the ability of PKF to efficiently deal with nonlinear and mixed dynamics.

\section{Conclusion}   \label{sec:concl}

An approach for functional sets with typed symbols is introduced in this work. An explicit distinction between syntax and semantics helps formalize the management of dependencies, characterize sources of conservatism and analyze the impact of evaluation strategies (inner first vs. outer first/lazy/call-by-need). Image-sets with typed symbols generalize several set-representations like zonotopes and polynomial zonotopes to mixed domains, as exemplified with mixed polynotopes. The combination of polynomial functions with interval, signed and boolean symbolic variables through simple rewriting rules makes it possible to gather in a single compact and efficient data structure the description of non-convex and non-connected sets which would usually require costly bissection/splitting strategies to be propagated. The mixed-encoding of intervals proposed in this context allows to tune the granularity level of the discrete part of the description and, meanwhile, control the combinatorial complexity through the use of reduction operators. In addition, the traceability of uniquely identified typed symbols paves the way for advanced mixed sensitivity analysis and symbolic abstraction techniques.
The reachability results show the relevance of the proposed approach to deal with the verification and synthesis of Cyber-Physical Systems (CPS). 
Based on modeling tools for nonlinear hybrid systems, a compositional implementation of advanced reachability and filtering algorithms is made possible by simply using operator overloading. This has been exemplified with the proposed Polynotopic Kalman Filter (PKF) which paves the way to advanced hybrid nonlinear filtering techniques preserving inclusion.
Moreover, several bridges with random variables and stochastic filtering have been outlined as well as bridges with functional programming and object oriented paradigms, robust (and interpretable?) 
artificial intelligence \cite{Mirman2018,Gehr2018} with the neural network activation function ReLU, sensitivity analysis, and symbolic abstractions of hybrid systems. 
Much remains to be done to exploit these connections.

\begin{ack}                      
	The author would like to thank Prof. Ali Zolghadri for many insightful discussions during this research work and for his careful reading of the manuscript.   
\end{ack}

\bibliographystyle{plain} 
\bibliography{pkf_rev2Long_V304_archive}

\appendix

\end{document}